\RequirePackage{lineno} 
\documentclass[12pt]{article}

\usepackage{graphicx}
\usepackage{mathptmx}      % use Times fonts if available on your TeX system

\usepackage[onehalfspacing]{setspace}
\usepackage[english]{babel}
\usepackage{amssymb,amsmath,amsfonts}
\usepackage{natbib}
\usepackage{hyperref}
\usepackage{bm, bbm}
\usepackage{lineno}
\usepackage{geometry}
\usepackage{graphicx,color}
\usepackage{listings}
\usepackage{caption, subcaption}
\usepackage[utf8]{inputenc}

\newcommand{\data}{x}
\newcommand{\dataObs}{x^0}
\newcommand{\datasim}{x^{\mbox{sim}}}
\newcommand{\distance}{d}
\newcommand{\Model}{\mathcal{M}}
\newcommand{\parameter}{\theta}

\newcommand{\prior}{\pi}
\newcommand{\statistics}{S}
\newcommand{\estparameter}{\hat{\theta}}

\newcommand{\lossfunc}{\mathcal{L}}
\newcommand{\threshold}{\gamma}
\newcommand{\R}{\mathbb{R}}
\newcommand{\E}{\mathbb{E}}
\DeclareMathOperator*{\argmin}{arg\,min}

\newcommand{\blind}{0}
\begin{document}
%%%%%%%%%%%%%%%%%%%%%%%%%%%%%%%%%%%%%%%%%%%%%%%%%%%%%%%%%%%%%%%%%%%%%%%%%%%%%%

\if0\blind
{

\title{\bf Distance-learning For Approximate Bayesian Computation To Model a Volcanic Eruption} 

\author{Lorenzo Pacchiardi$^1$,
Pierre Kunzli$^2$\thanks{Contributed equally as the first author}, Marcel Schoengens$^3$, Bastien Chopard$^2$,\\ Ritabrata Dutta$^4$\thanks{Corresponding author: Ritabrata.Dutta@warwick.ac.uk}\\ 
{\em \small $^1$Department of Statistics, University of Oxford, UK}\\
{\em \small $^2$Computer Science Department, University of Geneva, Switzerland}\\
{\em \small $^3$Six Group AG, Zurich, Switzerland}\\
{\em \small $^4$Department of Statistics, Warwick University, UK}\\ }

\maketitle } \fi

\if1\blind
{
  \bigskip
  \bigskip
  \bigskip
  \begin{center}
    {\LARGE\bf Title}
\end{center}
  \medskip
} \fi

\bigskip
Approximate Bayesian computation (ABC) provides us with a way to infer parameters of models, for which the likelihood function is not available, from an observation. Using ABC, which depends on many simulations from the considered model, we develop an inferential framework to learn parameters of a stochastic numerical simulator of volcanic eruption. Moreover, the model itself is parallelized using Message Passing Interface (MPI). Thus, we develop a nested-parallelized MPI communicator to handle the expensive numerical model with ABC algorithms. ABC usually relies on summary statistics of the data in order to measure the discrepancy model output and observation. However, informative summary statistics cannot be found for the considered model. We therefore develop a technique to learn a distance between model outputs based on deep metric-learning. We use this framework to learn the plume characteristics (eg. initial plume velocity) of the volcanic eruption from the tephra deposits collected by field-work associated with the 2450 BP Pululagua (Ecuador) volcanic eruption.\\
\vspace{1ex}\noindent%not present in the title

\section{Introduction}
\label{sec:Intro}
In recent centuries, developments in science and technology have allowed us to explore the expanding universe, discover unknown particles and find out how and why a society interacts and reacts. To explain the fascinating phenomena of nature, natural scientists develop complex `\emph{mechanistic models}' of stochastic nature, though the likelihood functions for these models are not easily tractable or available. Hence the hard question is how to choose the best model or how to calibrate these models given the data. The generation of data, given a configuration of parameters for these mechanistic models, is relatively easy (although computationally expensive). 

The main bottleneck of the inference for mechanistic models is the intractability of the likelihood function of the data generating process. Widely applicable frequentist or Bayesian inferential techniques cannot be directly applied for these models due to the absence of the likelihood function. To deal with models where likelihood calculations fail, other techniques have been developed which are collectively referred to as likelihood-free inference or \emph{approximate Bayesian computation (ABC)} \citep{lintusaari2017fundamentals}. 
In a nutshell, ABC algorithms are able to sample from an approximate posterior distribution of the parameters by finding values which yield simulated `fake' data resembling the observed data to a sufficient degree. To quantify the resemblance, we need to find a discrepancy measure between the simulated and observed dataset. 
The two main difficulties in application of ABC methods are the choice of the `discrepancy measure' and performing inference for `computationally expensive' models. In this manuscript, we use `metric-learning' \citep{suarez2018tutorial} and `deep metric-learning' (see for instance \citep{ge2018deep}) to learn a discrepancy measure between datasets and develop a `nested-parallelization' scheme based on the message passing interface (MPI) \citep{MPIForum} to deal with the expensive mechanistic models. 

We use ABC in conjunction with `deep metric-learning' and `nested-parallelization' to calibrate an expensive geo-scientific model of volcanic eruption using ABC. 
 By looking at the
volcanic depositions obtained from field work, geologists usually want to know
the plume characteristics (velocity profile, radius profile, total height etc.) of a volcanic eruption. To do so, here we consider a numerical model of volcanic eruption suggested in \cite{kunzli2016parallel} and infer the plume characteristics from field observations using ABC. 
 As a result of applying ABC to this
context, not only we can automatically and efficiently estimate the model
parameters, but we can also perform parameter uncertainty quantification in a
rigorous manner. This also provides us with a way to validate the numerical model. We illustrate the performance of the developed inferential scheme for a simulated dataset and 
a real data collected from 
field observation of ground tephra depositions at 72 ground locations associated with the 2450 BP Pululagua (Ecuador) volcanic eruption \citep{volentik2010modeling}.

In Sections~\ref{sec:VolErModel} and \ref{sec:likfinf}, we describe the volcanic eruption model and approximate Bayesian computation correspondingly. The `metric-learning and `deep metric-learning' for ABC are described in Section~\ref{sec:distlearn}, while the nested-parallelization' scheme is detailed in Section~\ref{sec:nestparal}. We apply the developed inferential scheme to learn plume characteristics for a simulated and a real volcanic eruption dataset in Section~\ref{sec:result}, concluding in Section~\ref{sec:conclusion}.

\section{Model for volcanic eruption}
\label{sec:VolErModel} 
\begin{figure}
\begin{center}
\includegraphics[width=\textwidth]{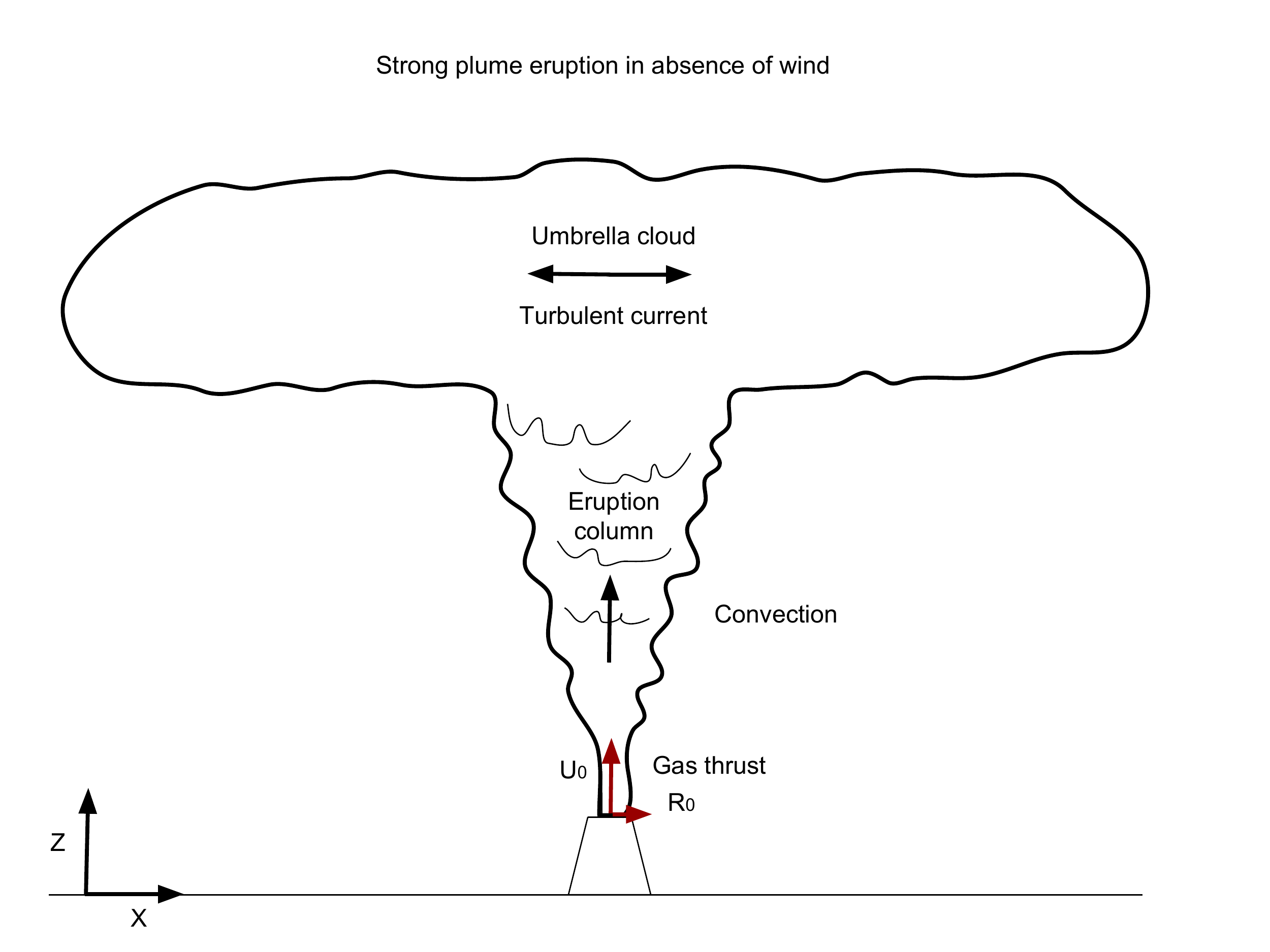}
\caption{\textbf{Strong Plume:} Representation of the main features of a strong plume.}
\label{fig:strongplume}
\end{center}
\end{figure}
During explosive volcanic eruptions, a hot mixture of particles and volcanic gases is typically ejected with an initial density several times larger than the atmosphere, and rises due to momentum. As the ejected material enters ambient air, the mixture density drastically decreases and an eruptive plume starts rising due to buoyancy. In the absence of wind (under the assumption of velocity of plume being much larger than the horizontal wind velocity), the volcanic plume buoyantly rises up to the neutral buoyancy level where it starts spreading laterally as a gravity current (umbrella cloud) [Figure~\ref{fig:strongplume}]. 

The physics of plume dynamics is described through volcanic ash transport and dispersal models \citep{degruyter_improving_2012}, which typically describe particle motion via a turbulent velocity field. Particles are advected inside this field from the moment they leave the vent of the volcano until they deposit on the ground. Several techniques exist to simulate particles in an advection field such as finite difference Eulerian, Lagrangian-puff or pure Lagrangian techniques. In this manuscript, we consider a model developed using a new numerically stable and easily parallelizable simulation tool called TETRAS (TEphra TRAnsport Simulator) based on a hybrid Eulerian-Lagrangian model developed in \cite{kunzli2016parallel}. The authors also provide a parallelization scheme on a distributed memory using MPI architecture since these models are computationally intensive to simulate. It also allows the computation of particle atmospheric concentration or ground mass load at any given time.

The main model parameters include the initial plume velocity $U_0$, radius of the plume at the vent of volcano $R_0$, initial temperature $T_0$ and the initial mass fraction of exsolved volatiles $n_0$. 
Depending on their size and density, particles are transported upward by the volcanic plume and, if sufficiently small, might be entrained within the umbrella cloud and sediment according to their terminal velocity. In particular, information on the Total Grain-Size Distribution (TGSD), namely the size distribution of particles injected into the atmosphere, combined with particle density is necessary to initialize the model. 
Turbulent effects, which play an important role in this model, are represented through a diffusion process. Different diffusion coefficients are applied in the atmosphere ($D_a$) and in the plume ($D_p$), the latter coefficient being usually several times larger. In the numerical model, the diffusion process is simulated by applying a random walk on particles (a random velocity is chosen for each particle at each iteration). The norm of the velocity and its direction are respectively chosen from a Gaussian distribution with given variance and a uniform distribution, to simulate a diffusion process with the prescribed diffusion coefficient. This procedure makes the model inherently stochastic. We refer the reader to~\cite{kunzli2016parallel} for more explanation on the mathematical model.

\begin{figure}[htbp]
\begin{center}
\begin{subfigure}{0.45\textwidth}
\begin{center}
\includegraphics[width=\textwidth]{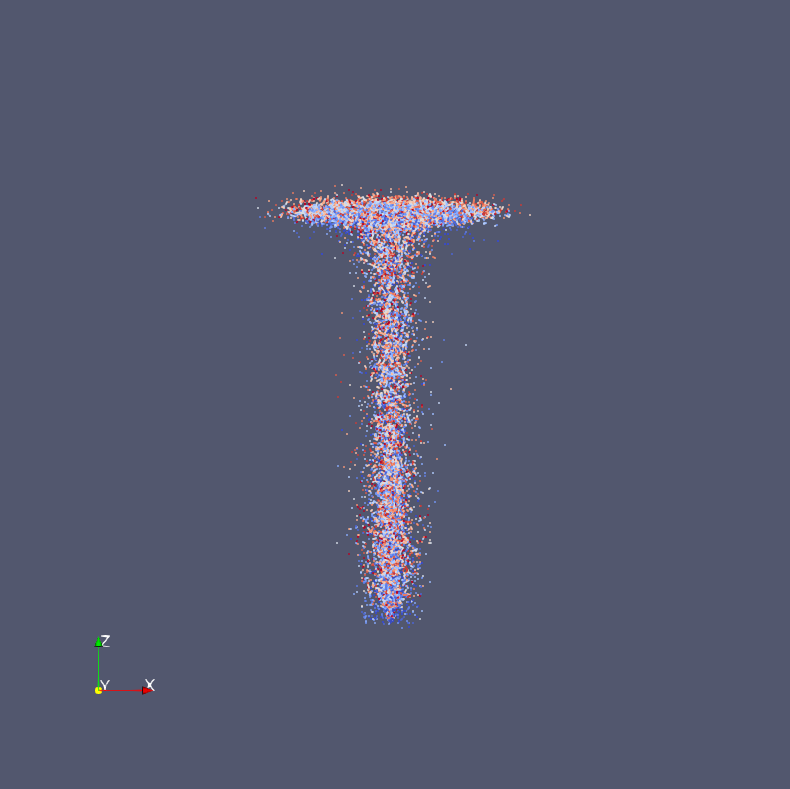}
\caption{$t=330$ sec.}
\end{center}
\end{subfigure}
\hspace*{\fill}
\begin{subfigure}{0.45\textwidth}
\begin{center}
\includegraphics[width=\textwidth]{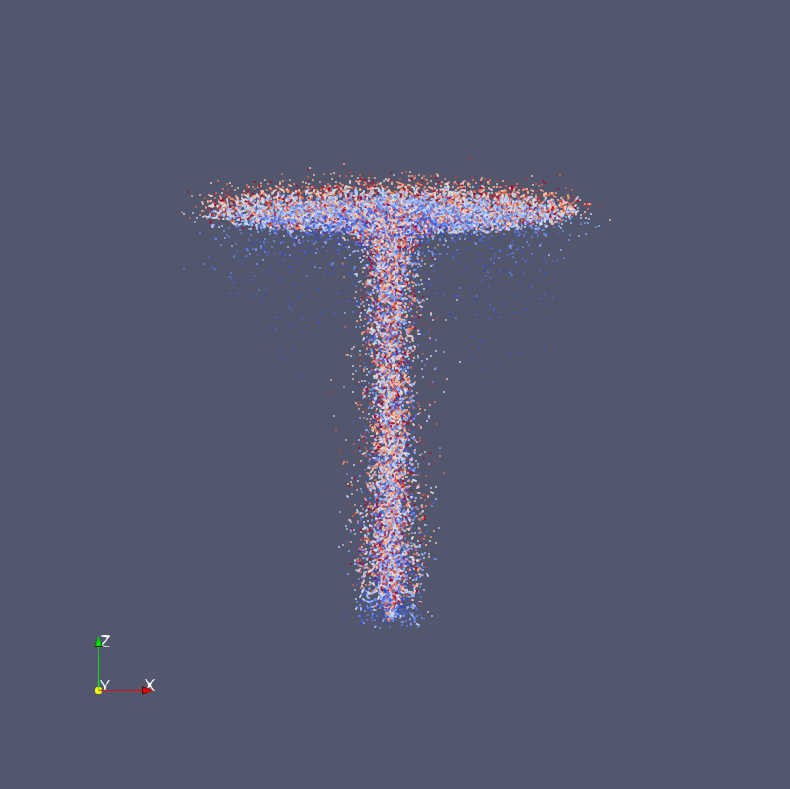}
\caption{$t=500$ sec.}
\end{center}
\end{subfigure}
\begin{subfigure}{0.45\textwidth}
\begin{center}
\includegraphics[width=\textwidth]{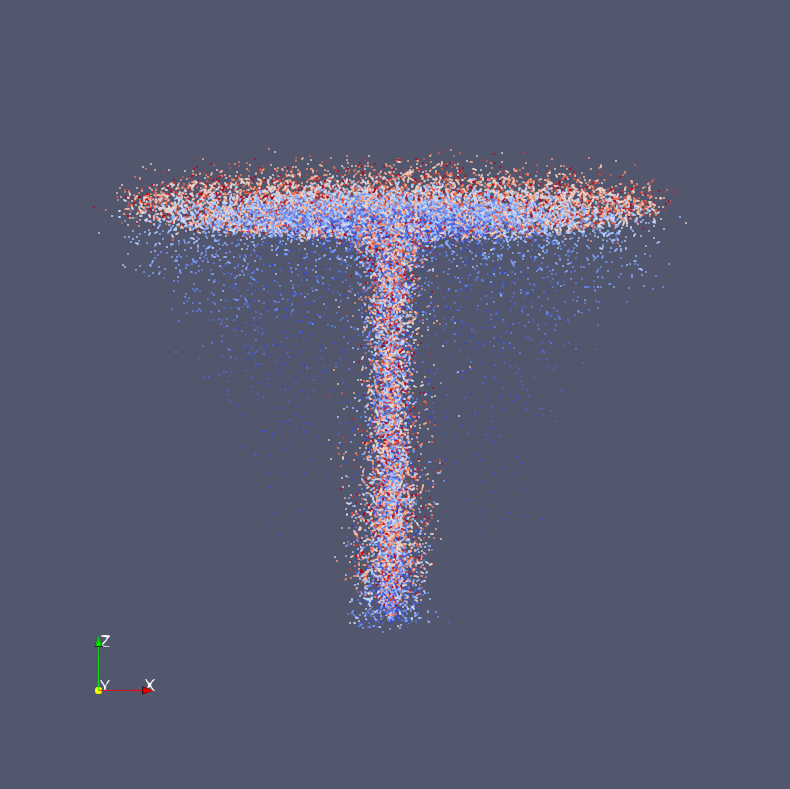}
\caption{$t=700$ sec.}
\end{center}
\end{subfigure}
\hspace*{\fill}
\begin{subfigure}{0.45\textwidth}
\begin{center}
\includegraphics[width=\textwidth]{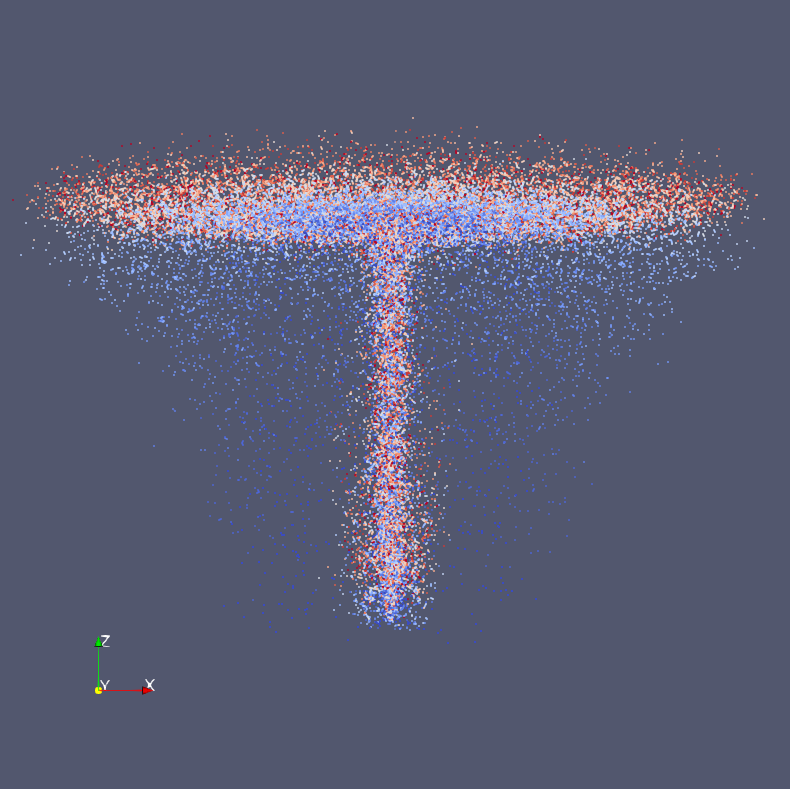}
\caption{$t=930$ sec.}
\end{center}
\end{subfigure}
\caption{\textbf{Simulation of a volcanic plume:} A volcanic plume is simulated by the numerical model over 15 minutes with parameters $U_0 = 250 [{m\over s}]$ and $R_0 = 50 [m]$. Colors represent size of particles. Grain size ranges from $\phi=-7$ (blue) to $\phi=10$ (red), $\phi=-\log[2](d)$ where $d$ is the diameter of the particle in mm.}
\label{fig:plumesimulation}
\end{center}
\end{figure}

The model forward-simulates a volcanic eruption by injecting particles at the vent into the domain and waiting for particles to either deposit on the ground or leave the domain. We consider the ground deposits of the volcanic plumes as the observed output of the volcanic eruption model, which can be measured through geo-scientific field work. A simulation of volcanic plumes by the model is illustrated in Figure~\ref{fig:plumesimulation} over 15 minutes.
Further, we will consider the field observation of ground tephra depositions at 72 ground locations associated with the 2450 BP Pululagua (Ecuador) volcano dataset \citep{volentik2010modeling}. Assuming the center of the volcano located at the same position as 2450 BP Pululagua (Ecuador) volcano, we consider the ground deposits at those 72 ground locations as the output of the volcanic eruption model.  

For the purpose of the present
study, the model $\Model$ is parametrized in terms of two of the parameters
introduced above, namely the initial plume velocity $U_0$ and the radius of
the plume at the vent of volcano $R_0$, collectively defined as $\parameter = (U_0, R_0)$.
In this work, we assume the initial temperature, the initial mass fraction of exsolved volatiles and 
the diffusion coefficients to be known. However, the model could also be parametrized in terms
of those parameters. 

Initial temperature $T_0$ and initial mass fraction of exsolved volatiles $n_0$
have an influence on plume height. They where chosen in a previous work \citep{kunzli2016parallel}
to produce a given plume height and have been kept constant as acceptable value
for this work. Their values are respectively $T_0 = 1256[K]$ and $n_0 \approx 0.01$.
Diffusion in atmosphere $D_a$ and diffusion in plume $D_p$ have been empirically chosen as
$D_a = 300\ m^2/s$ and $D_p = 1500\ m^2/s$.

By assuming that $\Model$ is true, then we can simulate the deposition of tephra at
those 72 locations, denoted as $\datasim$. If we have observed dataset
$\dataObs$, can we quantify the uncertainty or estimate the model parameters? As
the model described above \citep{kunzli2016parallel}, is stochastic in nature,
the observed data set could have been simulated using whole spectrum of values
for $\parameter$ with different likelihood. To quantify this stochastic
uncertainty in the parameters simulating the observed data set, we develop a
likelihood-free approximate Bayesian inference scheme in
Section~\ref{sec:likfinf}.

\section{Likelihood free inference}
\label{sec:likfinf}
We can quantify the  uncertainty of the parameter $\parameter$ by 
its posterior distribution $p(\parameter|\data)$ given the observed dataset $\data = \dataObs$. The posterior distribution is obtained by Bayes' theorem as
$p(\parameter|\dataObs) = \frac{\prior(\parameter)
 p(\dataObs|\parameter)}{m(\dataObs)},
$
where $\prior(\parameter)$, $p(\dataObs|\parameter)$ and $m(\dataObs) = \int\prior(\parameter)p(\dataObs|\parameter)d\parameter$ are, correspondingly, the prior distribution on the parameter $\parameter$, the likelihood function, and the marginal likelihood.
If the likelihood function could be evaluated, at least up to a normalizing constant, then the posterior distribution could be approximated by drawing a representative sample of parameter values from it using (Markov chain) Monte Carlo sampling schemes \citep{Robert2005}. Unfortunately,  the likelihood function induced by the volcanic eruption model is analytically intractable. 
In this setting, approximate Bayesian computation (ABC)  \citep{lintusaari2017fundamentals} offers a way to sample from an approximate posterior distribution and opens up the possibility of sound statistical inference on the parameter $\parameter$. In this paper we only focus on parameter estimation/calibration and
uncertainty quantification but we stress that ABC easily allows us to perform parameter hypothesis testing and model selection as well.

\subsection{Approximate Bayesian computation (ABC)}
\label{sec:abc}
\begin{figure}
	\begin{center}
		\includegraphics[width=0.7\textwidth]{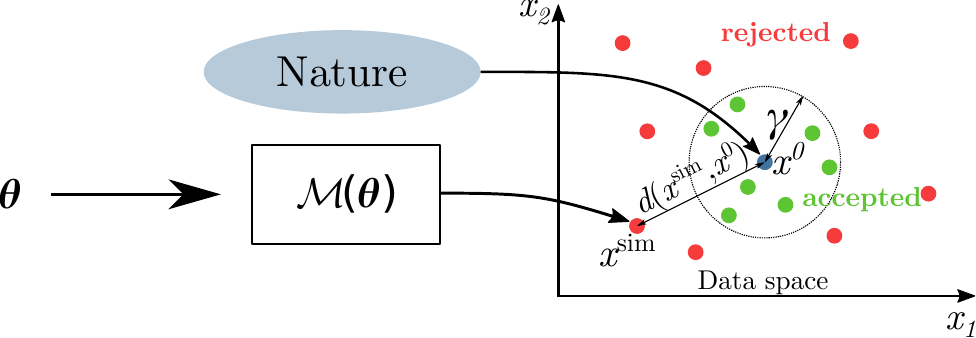}
		\caption{\textbf{ABC rejection sampling:} having observed data $ \dataObs $ provided by nature (the blue dot), we sample parameter values and generate observations through the simulator, that are then accepted (green) or rejected (red) according to their distance from the observation. The visualization is in a 2-dimensional data space.}
		\label{fig:ABC}
	\end{center}
\end{figure}

The fundamental ABC rejection sampling scheme iterates the following steps: 

\begin{enumerate}
	\item Draw $ \parameter $ from the prior $ \pi(\parameter)$.
	\item Simulate a synthetic dataset $\datasim$ from the simulator-based model $\Model(\parameter)$.
	\item Accept the parameter value $\parameter$ if $ \distance(\datasim,\dataObs) < \threshold $. Otherwise, reject $ \parameter $.
\end{enumerate}
See Figure \ref{fig:ABC} for a visualization of the above algorithm. 

Here, the metric on the dataspace $\distance(\datasim,\dataObs)$ measures the closeness between $\datasim$ and $\dataObs$. The accepted $(\parameter,\datasim)$ pairs are thus jointly sampled from a distribution proportional to $\pi(\parameter)p_{\distance,\threshold}(\dataObs|\parameter)$, where $p_{\distance, \threshold}(\dataObs|\parameter)$ is an approximation to the likelihood function $p(\dataObs|\parameter)$:

\begin{equation}\label{Eq:ABC}
p_{\distance, \threshold}(\dataObs|\parameter) = \int p(\datasim|\parameter) \mathbb{K}_{\threshold}(\distance(\datasim,\dataObs))  d\datasim, 
\end{equation}
where $\mathbb{K}_{\threshold}(\distance(\datasim,\dataObs)) $ is in this case a probability density function proportional to $\mathbbm{1}{(\distance(\datasim,\dataObs)<\threshold)}$\footnote{$\mathbbm{1}(\cdot)$ is used as an indicator function. }. Besides this choice for $\mathbb{K}_{\threshold}(\distance(\datasim,\dataObs))$, that has been exploited in several ABC algorithms (for instance \cite{beaumont:2010, drovandi2011estimation, del:2012, lenormand2013adaptive}), ABC algorithms relying on different choices exist, for instance being proportional to $\exp(-\distance(\datasim,\dataObs)/\threshold)$ in simulated-annealing ABC (SABC) \citep{Albert_2015}. In general, $\mathbb{K}_{\threshold}(\cdot )$ needs to be a probability density function with a large concentration of mass near 0, in which the parameter $ \threshold $ denotes the amount of concentration (the smaller $ \threshold $, the more concentrated the density is). This guarantees that, in principle, the above approximate likelihood converges to the true one when $ \threshold \to 0 $. Of course, decreasing the threshold increases the computational cost, as less simulations will be accepted.

More advanced algorithms than the simple rejection scheme detailed above are possible, for instance ones based on Sequential Monte Carlo \citep{del:2012, lenormand2013adaptive}, in which various parameter-data pairs are considered at a time and are evolved over several generations, while $ \threshold $ is decreased towards $0$ at each generation to improve the approximation of the likelihood function, so that you are able to approximately sample from the true posterior distribution. Alternative statistical methods for calibrating models from observations exist; however, the ABC framework has the advantage of both being applicable to stochastic models and of providing the user with a rigorous uncertainty quantification. For instance, methods based on Gaussian Process (GP) emulation, and subsequent use of the emulators for calibration, have been proved to work well \citep{o2006bayesian}, but mostly for deterministic models. Note also that some efforts of combining the versatility of ABC with the computational savings of using GP emulation have started appearing in the literature, see for instance \cite{meeds2014gps}, in which a GP is used to emulate the simulator and the need for new model runs is determined according to the uncertainty of the emulator; however, this relies on an ad-hoc algorithm. See also \cite{wilkinson2014accelerating} and \cite{gutmann2016bayesian} for examples of using GPs to emulate respectively the likelihood function and the discrepancy function. Another possibility is the use of an Ensemble Kalman Filter approach \cite{iglesias2013ensemble} to get an estimate of model parameters from an observation, but this does not provide an estimate of the uncertainty.

For the inference of parameters of the volcanic eruption model, here we choose the adaptive population Monte Carlo approximate Bayesian computation (APMCABC) algorithm, proposed in \cite{lenormand2013adaptive}, based on its suitability to high performance computing systems \citep{dutta2017abcpyhpc}.
At the first step of this algorithm, $N_{\mbox{sample}}$-many parameter values are randomly drawn from the prior distribution and the value of $\threshold$ is decreased adaptively depending on the pseudo data simulated from the model using those randomly sampled parameter values. In the next step, we produce $N_{\mbox{sample}}$-many parameter values approximately distributed from the distribution $p_{\distance,\threshold}(\parameter|\dataObs)$, for the adapted $\threshold$ value from last step and again decrease the $\threshold$ depending on the new samples. This procedure is continued $N_{\mbox{step}}$ many times or until some stopping criterion is reached. We note that the adapted $\threshold$ values at each step  are strictly decreasing and converge to zero, therefore improving the approximation to the posterior distribution. 
We finally note that this algorithm is extremely suitable to parallelization, as at each step, we always need to run the same number of forward simulations from the model; therefore, we can simply use a number of samples equal to the available workers, or choose the number of workers to allocate according to the number of samples we want to use.

\subsection{Distance learning}
\label{sec:distlearn}

%{\color{red} Explain what we do when multiple datapoints in dataset .... classification accuracy \citep{gutmann2017likelihood}, wasserstein distance \citep{bernton2019approximate} .... but simulation of multiple data is expensive: Reviewer 1}
%We remark that the ABC framework can also be applied in the case in which many observations of the same phenomenon are given. A possible technique to keep information from all the observed datasets is performing ABC separately for each of them sequentially, by using as the posterior on each inference as prior for the next one. However, we do not go into details here, as we will consider a single observation at a time. 

Traditionally, distance between $\datasim$ and $\dataObs$ are defined by summing over Euclidean distances between all possible pairs composed of one simulated and one observed datapoint in the corresponding datasets. Recently, distances for ABC has also been defined through accuracy of possible classification of $\datasim$ and $\dataObs$ \citep{gutmann2017likelihood} or by Wasserstein distance \citep{bernton2019approximate}, under the assumption that the datapoints in each datasets are identical and independently distributed and they are present in a large number in both $\datasim$ and $\dataObs$. We notice here that we only have a one datapoint in the observed dataset for a volcanic eruption field study and also due to the very expensive simulation model we can only have few datapoints in the simulated dataset. Hence, here we concentrate on the definition of distances through Euclidean distance while we only have one datapoint in both $\datasim$ and $\dataObs$.

While performing ABC for inference, problems may arise in cases where the data $ \data $ is high-dimensional. In fact the number of simulations needed before you get close enough to the observation increases with the dimension of the data space.  
Therefore, a common practice in ABC literature is to define $\distance$ as Euclidean distance between a lower-dimensional summary statistics $S: \datasim \mapsto S(\datasim)$, so that $\distance(\datasim,\dataObs)<\threshold$ would be replaced by $$ \distance(S(\datasim),S(\dataObs))<\threshold $$ in Eq.~\eqref{Eq:ABC}, boiling down to obtaining an approximation to the following likelihood function: 

\begin{equation}\label{Eq:ABC_with_statistics}
p_{\distance, \threshold}(\statistics(\dataObs)|\parameter) = \int p(\datasim|\parameter) \mathbb{K}_{\threshold}(\distance(\statistics(\datasim),\statistics(\dataObs))<\threshold)  d\datasim, 
\end{equation}
so that now $(\parameter,\datasim)$ are jointly sampled from a distribution proportional to $\pi(\parameter)p_{\distance,\threshold}(\statistics(\dataObs)|\parameter)$ when performing ABC inference. 

Reducing the data to suitably chosen summary statistics may also yield more robust inference with respect to noise in the data. Moreover, if the statistics is sufficient, then the above modification provides us with a consistent posterior approximation \citep{didelot_likelihood-free_2011}, meaning that we are still guaranteed to converge to the true posterior in the limit $ \threshold \to 0 $. As sufficient summary statistics are not known for most of the complex models, the choice of summary statistics remains a problem \citep{csillery_approximate_2010} and they have been previously chosen in a problem-specific manner \citep{blum_comparative_2013, fearnhead_constructing_2012, gutmann2018likelihood}. For volcanic eruption model, $y$ can not be easily transformed into summary statistics $S(y)$ as there is a complex spatial dependence involved between the deposited tephra at each locations. 
Hence, here we consider two possible ways of learning a distance directly between two datasets $\data_1$ and $\data_2$ rather than between the extracted summary statistics. The first entails constructing a Mahalanobis distance,
\begin{equation}
\label{eq:mahadist}
\distance_M(\data_1,\data_2) = \sqrt{(\data_1 - \data_2)^TM(\data_1-\data_2)}
\end{equation}
where $M$ is a $d \times d$ positive semi-definite matrix.

The second approach uses instead a neural network to transform non-linearly the dataset in a new space; this is usually referred to as \textit{deep} metric-learning and is a well developed field in the computer vision literature \citep{ge2018deep}. The learned distance is the Euclidean distance between the learned embeddings: 
\begin{equation}
\label{eq:NN_dist}
\distance_{NN}(\data_1,\data_2) = ||g_w(\data_1) - g_w(\data_2) ||_2,
\end{equation}
where $g_w(\cdot)$ denotes the transformation applied by the network with weights $ w $ and $ ||\cdot ||_2 $ denotes the $ L^2 $ norm.

In both cases, we aim to learn a distance function between data pairs approximating, in the best possible way, the Euclidean distance between the pair of parameters that generated them. Using a very good approximation of the Euclidean distance between the pair of parameters would be highly beneficial for ABC, as in this way the algorithm would be able to accept a simulated parameter value if and only if it is actually close to the parameter value generating the observation. 

This intuition can be better explained by first considering a deterministic model for which the map $ \parameter \mapsto \data $ is bijective. In this case, it is theoretically possible to learn a distance in data space that is exactly the same as the distance in parameter space. 
It would be in fact enough to apply the inverse model to the data, getting the parameter values generating them, and then compute the distance between the latter (although we stress that even finding the inverse of a deterministic model may be infeasible in practice). Note that, in this setting, the true posterior distribution of the parameters given the observation is degenerate in a single point, which would be a Dirac delta function at the parameter value generating the observation itself. Therefore using the above learned distance would be optimal in the ABC inference scheme, as the accepted values of $ \parameter $ would actually concentrate around the parameter value generating the observation and, as $ \threshold \to 0 $, we would get back the Dirac delta function. 

For the case of a deterministic model with non-bijective map $ \parameter \mapsto \data $, the previous justification does not hold anymore, as a given observation could have been originated by more than a single parameter value. It is therefore theoretically impossible to build a distance function between a pair of data samples that has the same value as the Euclidean distance between the parameters generating them (say `true distance'). However, a reasonable model would generate a given observation for parameter values that are relatively close by, eg. constituting a closed (and relatively small) patch in parameter space. Therefore, excluding unlikely scenarios, we argue that the distance learning approach would still be able to provide meaningful information, as it would be able to find some approximation of the true distance.

Finally, for the more general case of a stochastic model, the same argument as before still holds. In this case, in fact, the map $ \parameter \mapsto \data $ is non-bijective again; also, due to random noise, two observations generated from the same parameter value are likely to be at a positive distance, according to the learned measure. However, we argue that finding the closest distance function to the true distance is still useful, as it captures the stochastic part in the data dependent on the parameters. In fact, we expect that two samples generated from the same parameter value are assigned smaller (even if non-zero) distance than two samples generated by far apart parameter values. 
These heuristic justification still lacks theoretical guarantees and rigor; however, this goes beyond the scope of this work, therefore we will leave the investigation of this aspect for future works. 	

	Finally, we note that for the two cases discussed above, learning of the distance function corresponds to learning a transformation of the data. This can be immediately seen for the neural network based distance, where we consider the Euclidean distance between the transformed data using the transformation $ \data \mapsto g_w(\data) $. In the Mahalanobis distance case, instead, it is sufficient to recall that for each positive semidefinite matrix $ M $ there exists a square matrix $L $ such that $M = L^T L$. Therefore, we can write Eq.\ref{eq:mahadist} in the following way: 
	\begin{equation}\label{}
		    d_M(\data_1,\data_2) = \sqrt{(\data_1-\data_2)^T L^T L (\data_1-\data_2)} = \sqrt{(L(\data_1-\data_2))^T L (\data_1-\data_2)} = \| L(\data_1-\data_2)\|_2, 
	\end{equation}
	from which it is clear that the above corresponds to learning the transformation $ \data \mapsto L x $ and then computing the Euclidean distance between the transformed data. 

	However, we stress that our focus is different from the usual approaches of learning summary statistics as described in Section~\ref{sec:summarylearning}; in fact, we are motivated directly by the distance measure between pair of samples while, to the best of our knowledge, summary statistics learning is usually unrelated to the distance measure; see for instance \cite{prangle2015summary} for a review. For this reason, distance learning techniques consider several samples at a time (pairs, triplets or, possible, even more), while the summary statistics learning techniques mostly consider separately each (parameter-data) sample (see for instance the linear regression technique by \cite{fearnhead_constructing_2012}).

\subsubsection{Learning the distance from the data}\label{Sec:learning_distance}

We now discuss practical ways to learn the matrix $ M $ and the weights of the network. 
Following the discussion in the previous section, we consider here the assumption that the geometry induced in data space by these distances should be similar to the geometry in the corresponding parameter space induced by Euclidean distance ($\distance_E$).

We proceed therefore in the following way: we simulate a set of $n$ datasets $\lbrace \data_1, \ldots, \data_n \rbrace$ from $n$  parameters $\lbrace \parameter_1, \ldots, \parameter_n \rbrace$ correspondingly. In order to capture the information about the geometry of the parameter space, we define two complementary sets of  pairwise  similarity  constraints
$\mathbb{S} = \lbrace(\data_i, \data_j)| \data_i \mbox{ and } \data_j \mbox{ are similar}\rbrace$
and  dissimilarity constraints
$\mathbb{D} = \lbrace(\data_i, \data_j)| \data_i \mbox{ and } \data_j \mbox{ are dissimilar}\rbrace$, where $\data_i \mbox{ and } \data_j$ are considered similar if $\distance_E(\parameter_i,\parameter_j) < \epsilon$ and dissimilar otherwise, for some $\epsilon>0$.

\paragraph{Learning Mahalanobis distance:} We describe now how to learn a Mahalanobis distance, as in Equation~\ref{eq:mahadist} under the above similarity and dissimilarity constraints. This setup falls under a well-developed field of research in metric-learning \citep{suarez2018tutorial}. Here, we consider a $l_1$-penalized log-determinant regularization on $M$ \citep{ravikumar2011high}, which reduces the above distance learning problem to a $l_1$-penalized log-det optimization problem to find $M$: 
\begin{equation}\label{Eq:SDML}
\mathop{min}_{M} \mbox{tr}(M_0^{-1}M) - \mbox{log det} M +\lambda \sum_{i \neq j}M_{ij} + \eta \sum_{i,j=1}^n \left(\data_i^TM\data_i - \data_i^TM\data_j\right) K_{ij} 
\end{equation}
such that $M\geq 0 $ (is a positive semidefinite matrix) and 
\begin{eqnarray}
K_{i,j} =    
 \begin{cases}
      +1, & \text{if}\ (\data_i, \data_j) \in \mathbb{S} \\
      -1, & \text{if}\ (\data_i, \data_j) \in \mathbb{D}.
    \end{cases}
\end{eqnarray}
In \ref{Eq:SDML}, the first term can pushes the matrix $ M $ to be similar to the inverse of $ M_0 $, that can be thought of as the inverse of the prior on the final $ M $. Moreover, the second term is a spectral regularization on the matrix, while the third one is enforcing sparsity in the off-diagonal elements of $ M $, with $\lambda$ controlling the amount of sparsity. Finally the fourth term is the one encoding the information coming from the similarity and dissimilarity sets; the trade-off between the latter and the previous terms is tuned by $ \eta $. This algorithm is called Sparse Distance metric-learning (SDML) \citep{qi2009efficient}. 

\paragraph{Deep metric-learning:} For the second approach, to learn the weights of the neural networks, here we consider the contrastive \citep{hadsell2006dimensionality} and triplet \citep{schroff2015facenet} losses defined on the same similarity/dissimilarity constraints as above. The learned distances will be called correspondingly contrastive loss distance and triplet loss distance. The contrastive loss considers all possible pairs of samples and penalizes a large embedding distance for similar samples while, for dissimilar ones, it penalizes them for being too close, and pushes them to be further apart than a fixed margin $ \alpha $. 
Specifically, we can write it in the following form:
\begin{equation}\label{eq:contrastive}
L = \frac{2}{n(n-1)} \sum_{i=1}^{n} \sum_{j=i+1}^{n}\left\{ y_{ij} \cdot ||g_w(\data_i) - g_w(\data_j) ||_2^2 + (1- y_{ij}) \cdot [\alpha - ||g_w(\data_i) - g_w(\data_j) ||_2]_+^2\right\}, 
\end{equation}
where $ [\cdot]_+ = \max(0,\cdot) $ and where $ y_{ij} = 1 \iff (\data_i, \data_j) \in \mathbb{S} $, $ y_{ij} = 0 \iff (\data_i, \data_j) \in \mathbb{D} $.

The triplet loss works instead on three samples at a time: an anchor, a positive, that is deemed similar to the anchor, and a negative, that is on the contrary dissimilar. Essentially, the loss pushes the network to find an embedding such that the distance between the anchor and the negative is larger than the one between the anchor and the positive plus a margin, that is defined a priori. By denoting $(\data_a^{(i)}, \data_p^{(i)} , \data_n^{(i)})$ the anchor, positive and negative of the \textit{i}-th triplet, and by denoting as $ N $ the number of all possible triplets built in this way, we can write the loss in the following way: 
\begin{equation}\label{eq:triplet}
L = \frac{1}{N} \sum_i^{N} \left[|| g_w(\data_a^{(i)}) - g_w(\data_p^{(i)})||_2^2 - || g_w(\data_a^{(i)}) - g_w(\data_n^{(i)})||_2^2 + \alpha\right]_+,
\end{equation}
where $\alpha \in \R$ denotes again the margin. We optimize this loss with stochastic gradient descent over the parameters of the network, by drawing random triplets. 
%Note that, as the number of triplets grows cubically with the number of samples, this is extremely inefficient when the number of samples is large. Also, as training proceeds, it becomes more and more likely that the random triplet is an "easy" one, i.e. for which $ || g_w(\data_a^{(i)}) - g_w(\data_n^{(i)})||_2^2 > || g_w(\data_a^{(i)}) - g_w(\data_p^{(i)})||_2^2 + \alpha $ and that the loss evaluated there is 0. Therefore, a more careful choice of the triplets may be necessary. However, in our case, when using the same $ n=400 $ training samples used to learn the Mahalanobis distance, and having quite a small data size (72, leading to a small number of network parameters), a better choice of the triplets did not bring any practical advantage. We will not discuss it further for this reason. 

While defining the similarity and dissimilarity constraints, $\epsilon$ was chosen to be the 10-th percentile of the pairwise distances between the $n$ parameters $\lbrace \parameter_1, \ldots, \parameter_n \rbrace$, for both SDML and deep metric-learning. 
To optimize SDML, we use an iterative optimization scheme from \citep{qi2009efficient} implemented in the \texttt{metric-learn} Python package \citep{metric-learn}, with $M_0$ chosen to be the sample covariance matrix, $\eta = 0.15$, $\lambda=0.01$ and $n=400$. 
For deep-metric-learning, we have used a 4-layers fully connected network, with 72 input neurons and 15 outputs, and with hidden layers of size 100, 80 and 40. We used ReLU non-linearity between the layers, $\alpha=1$ and Stochastic Gradient Descent for both losses, drawing random pairs or triplets. Note that the size of the embedding, 15 has been hand-tuned based on some pilot runs and sensitivity analysis. A more rigorous data-driven choice of size of embedding is left for future work. 
%in this case, was fixed quite arbitrarily; some experiments were conduced and small sensitivity of the results with respect to this was found. However, we leave the investigation of theoretical hints towards an optimal embedding size in future work.

We stress that the network we used is very small compared to the ones usually considered in computer vision applications, in which these techniques were firstly developed; also, another conceptual difference exists: in computer vision, the deep metric-learning techniques are used in a supervised setting, in which every image is assigned a label and similar pairs consist of images of the same class. Our case, instead, is what may be called a \textit{weakly}-supervised context, in which the only information we have is the similarity set. Note that, in the former case, $ (x_1, x_2), (x_2, x_3) \in \mathbb{S} \implies (x_1, x_3) \in \mathbb{S}$, while this is not true in the weakly-supervised case. 

\begin{table}[htbp]
	\centering
\begin{tabular}{c|c|c|c}
	\textbf{Loss} & \textbf{Number of epochs}  & \textbf{Batch size } & \textbf{Margin $ \alpha $}\\ 
	\hline 
	{Contrastive}	& 400 & 32 & 1\\
	{Triplet}	& 800 & 16 & 1\\ 
	{Semiautomatic NN}	& 400 & 2 & n.a.\\ 
\end{tabular} 
\caption{Settings for neural network training.}
\label{Tab:setting}
\end{table}
 
Please refer to Table \ref{Tab:setting} for the number of epochs and batch size used for deep metric-learning. At each epoch, we iterate over all samples and draw another random element, in the contrastive case, or a random positive and random negative in the triplet case. For the contrastive loss, as the similar pairs are fewer than the dissimilar ones, a random pair would be more probably dissimilar than similar. In order to enhance the training, we therefore sample with probability $ p=0.4 $ a similar sample to the considered one, and with remaining probability a dissimilar sample; in this way, the fraction of positive pairs on which the network is trained is larger than what it would be by naively sampling another random element. 

\begin{figure}[htbp]
\begin{center}	
\begin{subfigure}{0.18\textwidth}
\begin{center}
\includegraphics[width=\textwidth]{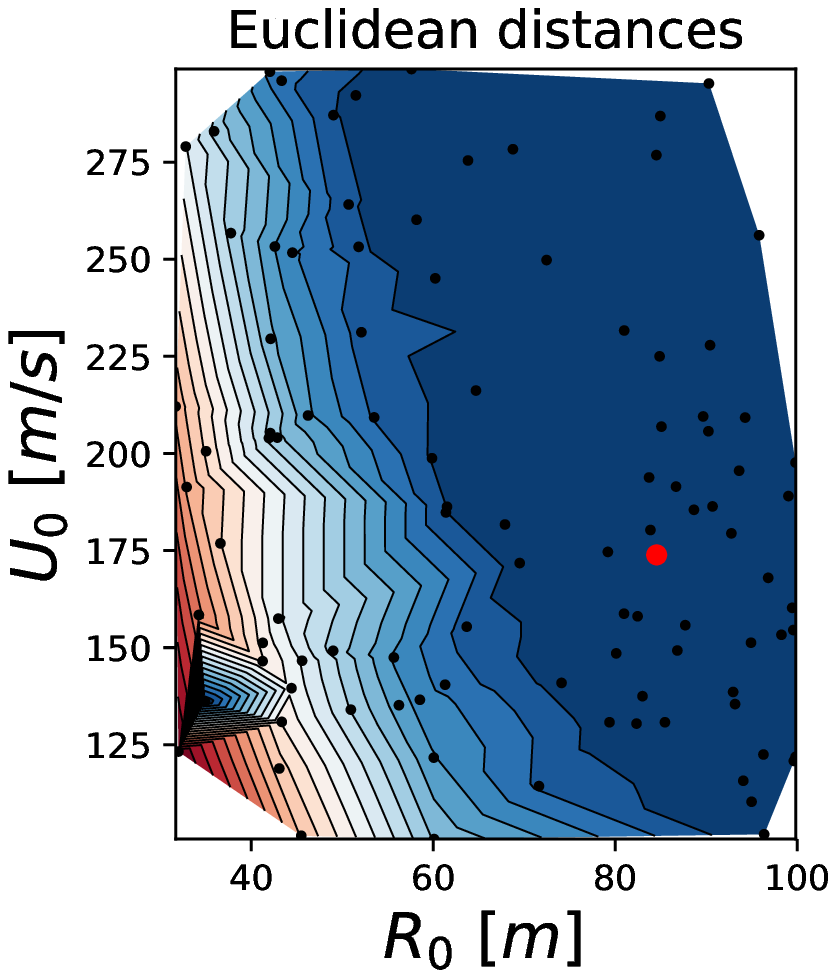}
%\caption{Euclidean}
\end{center}
\end{subfigure}
\hspace*{\fill}
\begin{subfigure}{0.18\textwidth}
\begin{center}
\includegraphics[width=\textwidth]{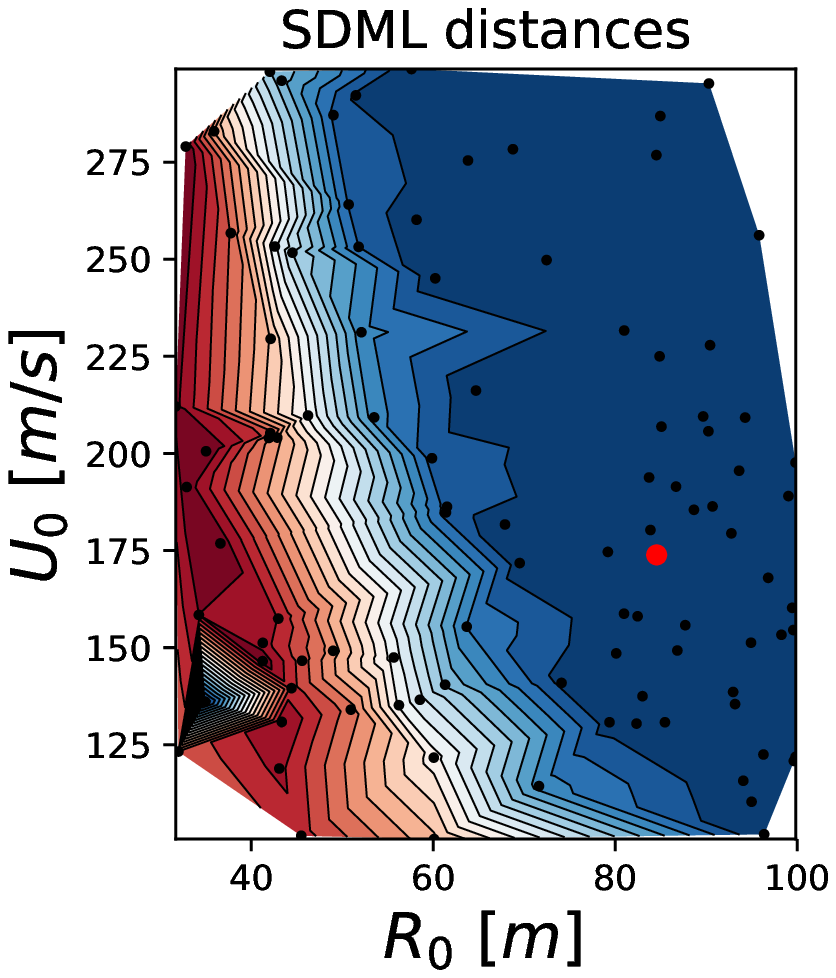}
%\caption{Distance Learned}
\end{center}
\end{subfigure}
\hspace*{\fill}
\begin{subfigure}{0.18\textwidth}
	\begin{center}
		\includegraphics[width=\textwidth]{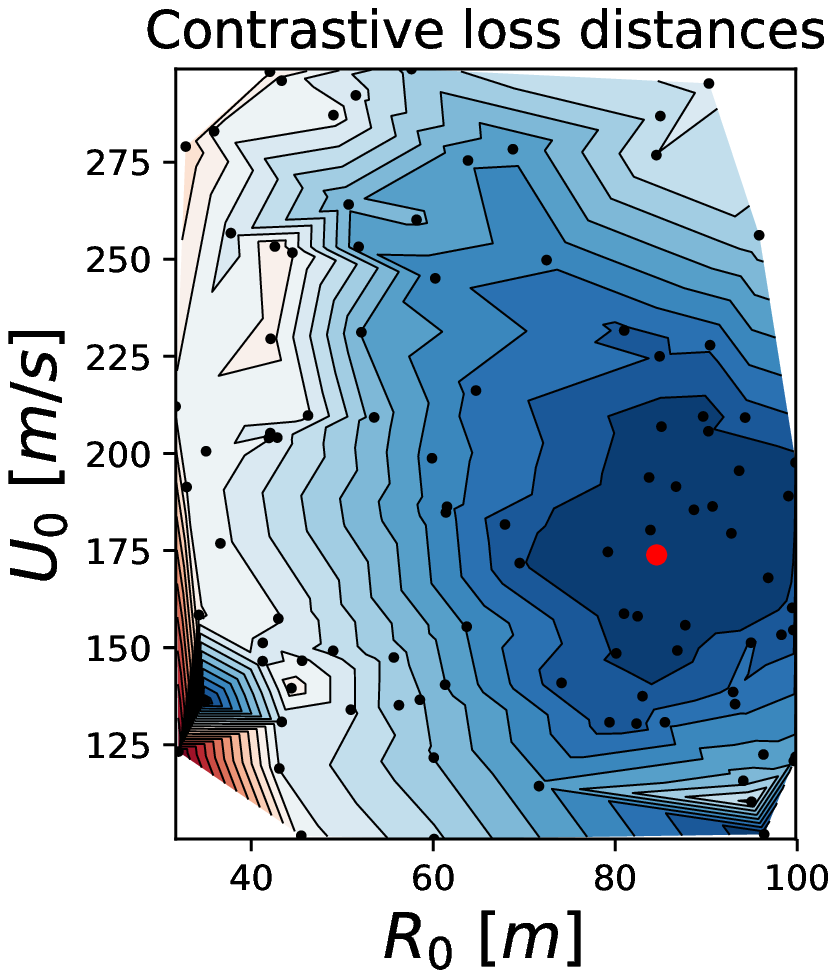}
		%		\caption{Distance Learned}
	\end{center}
\end{subfigure}
\hspace*{\fill}
\begin{subfigure}{0.18\textwidth}
	\begin{center}
		\includegraphics[width=\textwidth]{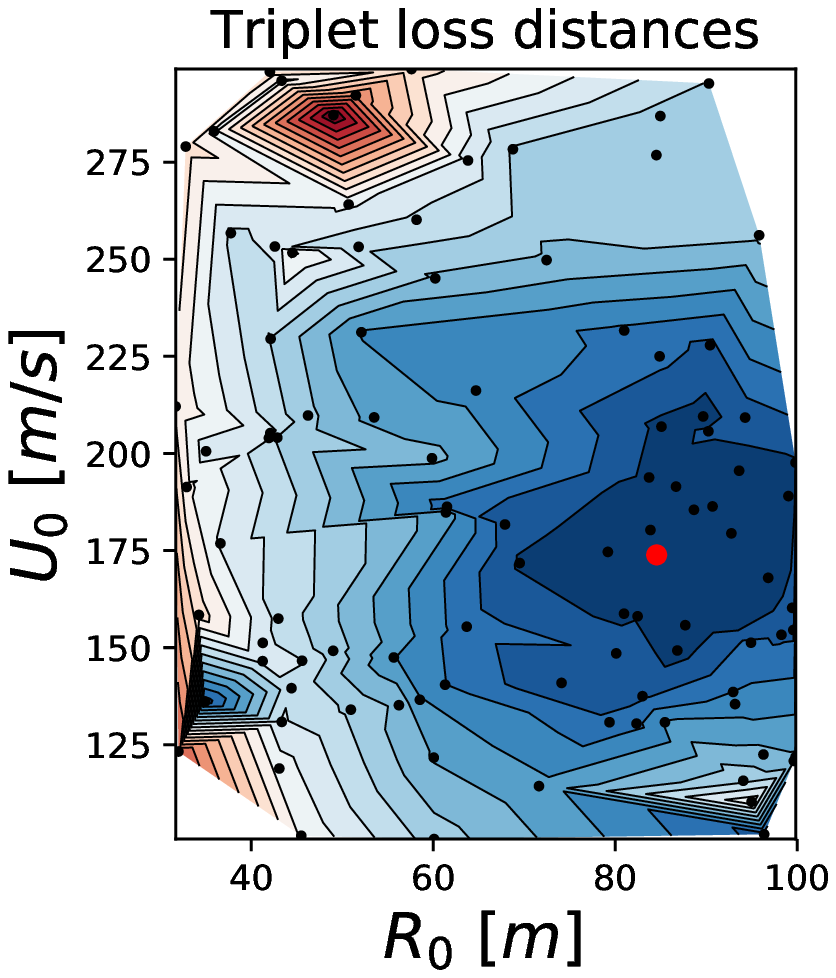}
%		\caption{Distance Learned}
	\end{center}
\end{subfigure}
\hspace*{\fill}
\begin{subfigure}{0.18\textwidth}
	\begin{center}
		\includegraphics[width=\textwidth]{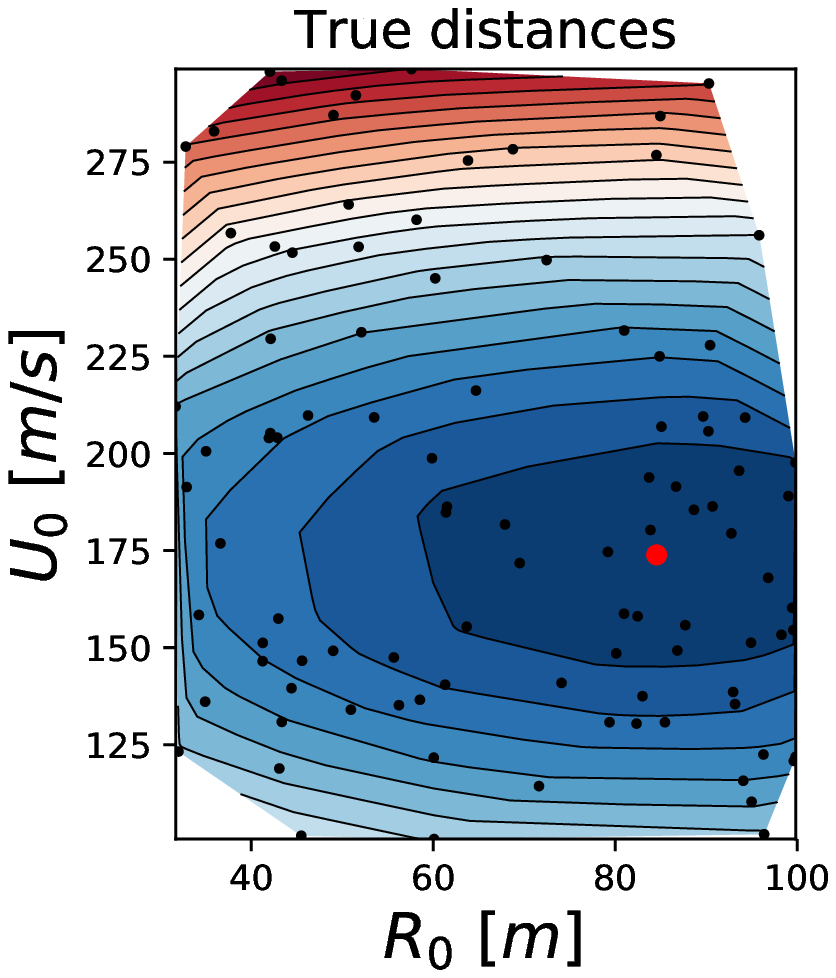}
%		\caption{Distance Learned}
	\end{center}
\end{subfigure}
\hspace*{\fill}
\begin{subfigure}{0.045\textwidth}
	\begin{center}
		\includegraphics[width=\textwidth]{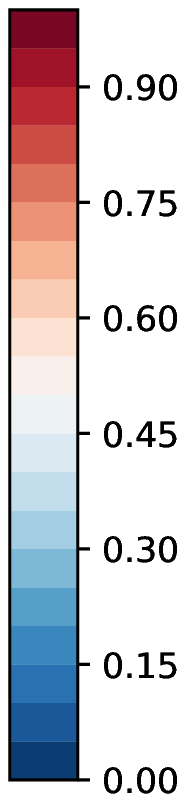}
		%		\caption{Distance Learned}
	\end{center}
\end{subfigure}
	\caption{\textbf{Comparison of metrics:} We compare Euclidean distance $\distance_E$ and the distances learned with the discussed techniques on the 300 elements of the training set, computing them between an observation point $\dataObs$ and the other 99 samples in the test set, taken to be a reference. $\dataObs$ was simulated using $U_0 = 173.87\ m/s,\ R_0=84.55\ m $ [Red point]. All the distances are scaled between [0,1]. The black small points denote the position of the reference data, from which the contour plot is obtained by triangulation. For reference, the Euclidean distance between the true value of the parameters is also reported. }
	\label{fig:comp_metric}
\end{center}
\end{figure}

In Figure~\ref{fig:comp_metric}, we compare the Euclidean distance between the parameters generating the datasets (`true distance') with the learned distance functions on the corresponding datasets, namely the Mahalanobis one with SDML algorithm and the contrastive and triplet loss distances; we also report the Euclidean distance between outputs of the model. The comparison is done in the following way: 400 parameters-simulation pairs have been generated, with parameters drawn uniformly on the interval $ (30,100)\ [m] $ for $ R_0 $ and $ (100,300)\ [m/s] $ for $ U_0 $. This dataset is split into a training one (with 300 samples) and a test one (with 100 samples). We learn the distances on the training set, and then compute all the distances between a chosen element in the training set $\dataObs$ and the 99 remaining samples in the same test set (`reference samples'); we then plot the learned distances in parameter space, by using the corresponding parameter value for each observation. $\dataObs$ was simulated using $\parameter_0 = (173.87\ m/s, 84.55\ m)$. 
We see that the minimum values of the distances are much more concentrated around the true parameter value for the contrastive and triplet loss distance in comparison to the SDML one and the Euclidean. Note also that the neural network based ones are able to partially reproduce the behavior of the distance between the true parameter values. 
However, it is not clear from this visualization which one between contrastive and triplet performs better. We therefore perform a more rigorous comparative study in the next Section, in order to find out which is the best between the two deep metric learning techniques and to evaluate different choices of $ \epsilon $.

As a side remark, we note that the number of possible pairs and triplets grows respectively quadratically and cubically with the number of training samples. This implies, for the triplet case, the number of triplets that is seen by the network during the training is smaller than the total number of them, for our chosen number of epochs. However, the network is still capable of reproducing the behavior of the true distance, and further training did not seem to produce any improvement. Of course, these techniques are extremely inefficient when the number of samples is large; techniques have been developed in order for the network to focus on hard pairs and triplets only. However, as our training size is quite small (300) we do not discuss these in detail here, and we refer to \cite{liu2019general} and \cite{hermans2017defense}.

\subsubsection{Comparison using Kullback-Leibler divergence}\label{Sec:KL}

In order to obtain a quantitative comparison between different distance learning techniques discussed and different choices of $ \epsilon $, we estimated the Kullback–Leibler (KL) divergence between a distribution induced by the learned distance function on parameter space (with respect to some reference point), and the distribution induced in the same way by the true distance.
Specifically, we first learn a distance with one of the methods described above on the train set, and then compute the distances of all reference samples in the test set with respect to another simulated dataset $ \dataObs $ corresponding to parameter $ \parameter_0 $ (the `observation'), as done in the previous Section. Then, after scaling the distances to $ [0,1] $ in the considered region, we consider a Gibbs density defined on the reference values of parameters $ \parameter_i $ to be $ p(\parameter_i) \propto e^{-\beta \cdot \distance(\data_i, \data_0)^2} $, where $ \distance(\data_i, \data_0) $ is the learned distance function; we assume that the above distribution exists on all the parameter space (neglecting the fact that the map from $ \parameter $ to $ \data $ is stochastic), but that we can evaluate it only on the reference points. We consider also the distribution defined by the true distance between reference values of parameters and the observation one: $ p^*(\parameter_i) \propto e^{-\beta \cdot  \distance_E(\parameter_i, \parameter_0)^2/c} $, where $ c $ is a constant rescaling the distance to $ [0,1] $ in the considered region. Now, as we can evaluate the learned distance only on the reference points, we estimate the KL divergence using an importance sampling approach, that is described below; we apply this technique on the same set of $ n=400 $ simulations with the same train-test split that we discussed above, the parameters of which were drawn independently according to a uniform on the interval $ (30,100)\ [m] $ for $ R_0 $ and $ (100,300)\ [m/s] $ for $ U_0 $.

Recall now the definition of the KL divergence: 

$$D_{KL}(P||P^*) = \int p(\theta) \log \left( \frac{p(\theta)}{p^*(\theta)} \right) d\theta= \int q(\theta) \frac{p(\theta)}{q(\theta)} \log \left( \frac{p(\theta)}{p^*(\theta)} \right) d\theta, $$
where we denoted as $q$ the density according to which the parameters are drawn (uniform in our case), and where $ P $ (respectively $ P^* $) denotes the distribution with density $ p $ ($ p^* $). As we do not know the normalization constants of the above densities, we need to estimate them from the data. We define therefore the unnormalized densities $p(\theta) = \tilde p(\theta) /Z$ and $p^*(\theta) = \tilde p^*(\theta) /Z^*$. Then, we can estimate the divergence by: 

$$\hat D_{KL}(P||P^*) = \sum_{i=1}^n w_i \cdot \log \left( \frac{\tilde p(\theta)/ \hat Z}{\tilde p^*(\theta)/ \hat Z^*} \right),  \quad w_i = \frac{p(\theta_i)/q(\theta_i)}{\sum_{j=1}^n p(\theta_j)/q(\theta_j)} = \frac{\tilde p(\theta_i)/q(\theta_i)}{\sum_{j=1}^n \tilde p(\theta_j)/q(\theta_j)}, \quad \theta_i \sim q. $$
where $\hat Z = \frac{1}{n} \sum_{i=1}^n \frac{\tilde p(\theta_i)}{q(\theta_i)}$ is a consistent estimator of $Z$, and similarly for $\hat Z^* $.

Overall, we are then left with the following consistent estimator: 

$$\hat D_{KL}(P||P^*) = \sum_{i=1}^n \frac{\tilde p(\theta_i)/q(\theta_i)}{\hat Z} \cdot \left( \log\frac{\hat Z^*}{\hat Z} + \beta\left(\frac{d_E(\theta_i) }{c}- d(\theta_i)\right)     \right) , $$
where we have used the explicit dependence of $p$ and $p^*$ on the distance function. 

In order to have better statistics for the performance of each distance learning technique, we perform Leave-One-Out cross validation on the test set: after having learned the distance on the training set, for each of the samples $ (\data_j, \parameter_j) $ in the test set in turn, we consider it as an observation point in the computation described above, while all the other elements in the test set are taken as the reference $\lbrace \parameter_1, \ldots, \parameter_n \rbrace$ and used to estimate the KL divergence. 
In this way, we are able to obtain a statistics of the estimated KL divergence on 100 realizations.

\begin{figure}
	\begin{center}
				\includegraphics[width=\textwidth]{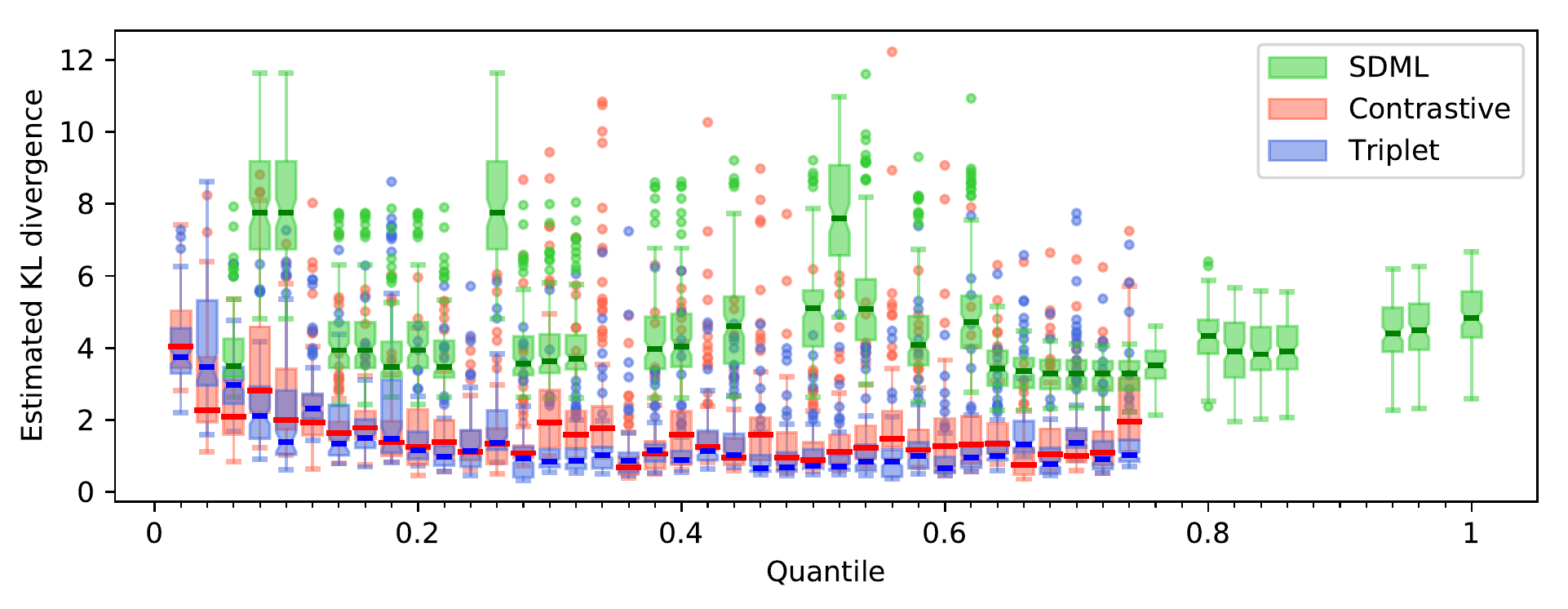}
				%				\caption{}	
		\caption{\textbf{Boxplot for the estimated KL divergence for different choices of the quantile defining the pairwise similarity sets.} Darker bars represent median over 100 realizations. The same train-test split and the same settings for the network as in Section~\ref{Sec:learning_distance} have been used.}
		\label{fig:epsilon_sensitivity}
	\end{center}
\end{figure}

We repeat this evaluation over the range of quantiles defining the similarity set over which the distances are learned. 
For each quantile value and technique we draw a boxplot, representing the spread of the histogram; the results can be found in Figure~\ref{fig:epsilon_sensitivity}. Recall that the KL divergence between two distributions is 0 if and only if they are the same, and can never be negative. 
Note that the SDML algorithm is quite unstable and was not able to converge for some of the $ \epsilon $ values. Also, the deep metric-learning techniques algorithms are not applicable for quantiles larger than $ 0.74 $, as in this case there is at least a training sample which is considered to be similar to all the other training samples, and the training routines are not designed to operate in this case. 

From the results, you can see that the triplet loss performs consistently better than the contrastive one, as the estimated KL divergence spans smaller values. 
Also, SDML is always worse than both of them and its performance does not show a clear trend with respect to the quantile. Regarding the deep metric-learning techniques, they capture more information with a quite large quantile value, i.e. in the case where a large fraction of all possible sample pairs are considered to be similar; this result is quite surprising. 
Finally, we note that the numerical value of the estimated KL divergence depends strongly on the choice of $ \beta  $, but the ranking between the different techniques remains the same; the results in Figure~\ref{fig:epsilon_sensitivity} was obtained with $ \beta =1 $. 

We choose as best distance learning technique the one which is able to achieve the smallest median of the KL histogram obtained over the possible 100 splits. 
Therefore, the best distance is found to be the triplet loss one trained by using the 60th percentile as threshold for defining the similarity set. 

\subsection{Semiautomatic summary statistics selection}
\label{sec:summarylearning}
We compare now the results of the distance learning approaches with the semiautomatic summary statistics learning schemes  \citep{fearnhead_constructing_2012, jiang2017learning}. In this approach, the parameter values are regressed using some function of the corresponding simulation outputs. Namely, you assume the following model: 

\begin{equation}\label{Eq:FPNN}
\parameter = \E(\parameter| \data) + \epsilon =  f_\beta(\data) + \epsilon,
\end{equation}
where $ \epsilon $ is a 0-mean noise and $ f_\beta(\data) $ is a function of data parametrized by $ \beta $. The authors of \cite{jiang2017learning} parametrize $ f_\beta(\cdot) $ by using a Neural Network. This regression approach was first introduced in \cite{fearnhead_constructing_2012} with a linearity assumpton on $ f_\beta $, reducing it to a simple linear regression. We focus here on on the neural network formulation as this was shown to outperform the linear regression by \cite{jiang2017learning}.

In practice, before performing ABC inference, the procedure amounts to the following steps:

\begin{itemize}
	\item We simulate $ N_{ss} $ data-parameter pairs $(\parameter_i, \data_i)_{i=1}^{N_{ss}}, \ \parameter_i \sim \pi(\theta), \ \data_i \sim \Model(\parameter_i)$
	\item We then fit the statistical model given by Eq.~\eqref{Eq:FPNN}.
	\item Finally, we fix $ \statistics(\cdot) = f_\beta(\cdot) $ in the chosen ABC inference algorithm (in Eq.~\eqref{Eq:ABC_with_statistics}).
\end{itemize}
	
In Theorem 3 of \cite{fearnhead_constructing_2012}, the authors provide a rationale for the above procedure; namely, they show that, by using $ \statistics(\dataObs) =  \E(\parameter| \dataObs)$  as summary statistics, the posterior mean of the ABC approximate posterior is the best possible estimator of the true parameter value with respect to the quadratic error loss, in the limit of $ \threshold \to 0$ in the ABC inference scheme (Eq.~\eqref{Eq:ABC_with_statistics}). Of course, the posterior mean with respect to the true posterior $  \E(\parameter| \dataObs) $ is not available, and hence the regression approach was proposed.

However, we highlight that the latter is only able to learn an approximation of the ``ideal" summary statistics, so that the theoretical justification is not conclusive. Therefore, we believe that directly focusing on learning the distance, as described in the previous Sections, may actually perform better than the regression approach in Eq.~\eqref{Eq:FPNN} in learning the best summary statistics. Intuitively, we think that the distance learning approach is able of modeling the correlations between different parameter values, as it relies on considering pairs or triplets of samples at a time. Again, we leave theoretical guarantees of the above intuitions for future work, and we simply rely on empirical studies  to make our point here.

We fit this model on the same set of datasets and simulation pairs that we used for the distance learning approach, using similar train and test split. As said above, we use a neural network to parametrize the function $ f_\beta(\cdot) $, and that was trained by stochastic gradient descent using the loss corresponding to the regression in Eq.\eqref{Eq:FPNN}: 
\begin{equation}\label{Eq:loss_fpnn}
\frac{1}{N} \sum_{i=1}^N   ||f_\beta(\data_i) - \parameter_i||_2^2.
\end{equation}

The neural network is composed of 4 fully connected layers, with 72 input neurons and 2 outputs, and with hidden layers of size 80, 40 and 15, with ReLU non-linearity. We remark that, in this case, the output dimension of the network (i.e. the dimension of the summary statistics) must match the number of parameters. Further details on the training settings may be found in Table~\ref{Tab:setting}.

\begin{figure}[htbp]
	\begin{center}
		\begin{subfigure}{0.2\textwidth}
			\begin{center}
				\includegraphics[width=\textwidth]{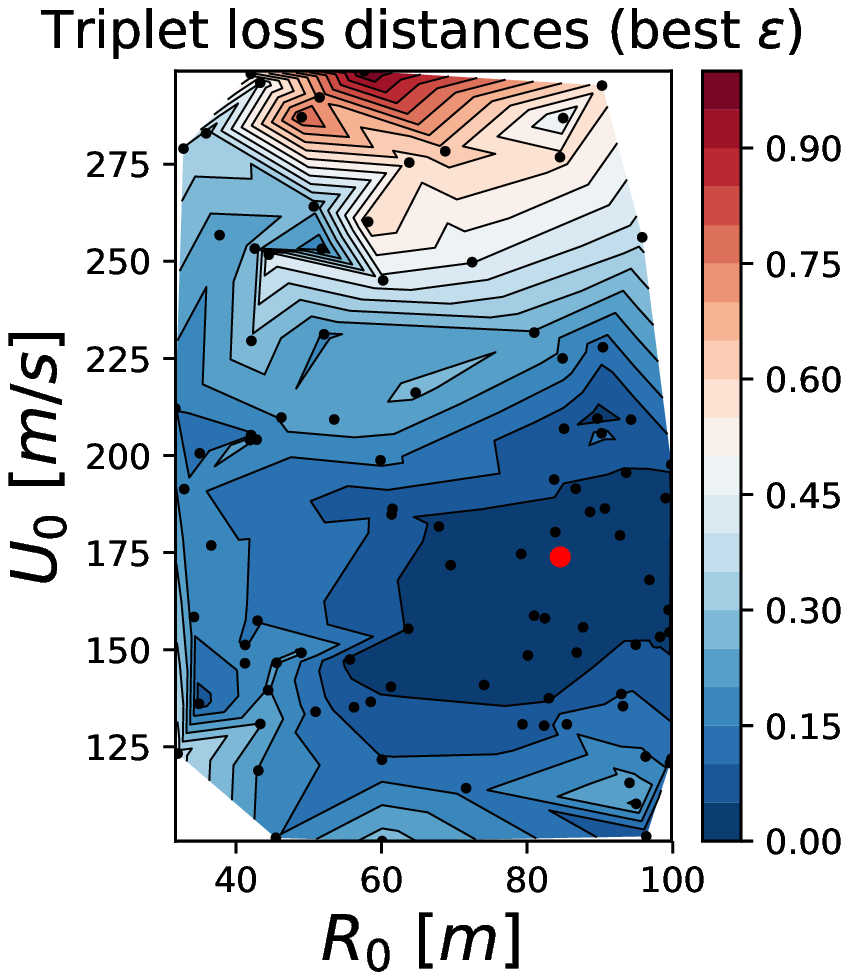}
%				\caption{}
			\end{center}
		\end{subfigure}~
		\begin{subfigure}{0.2\textwidth}
			\begin{center}
				\includegraphics[width=\textwidth]{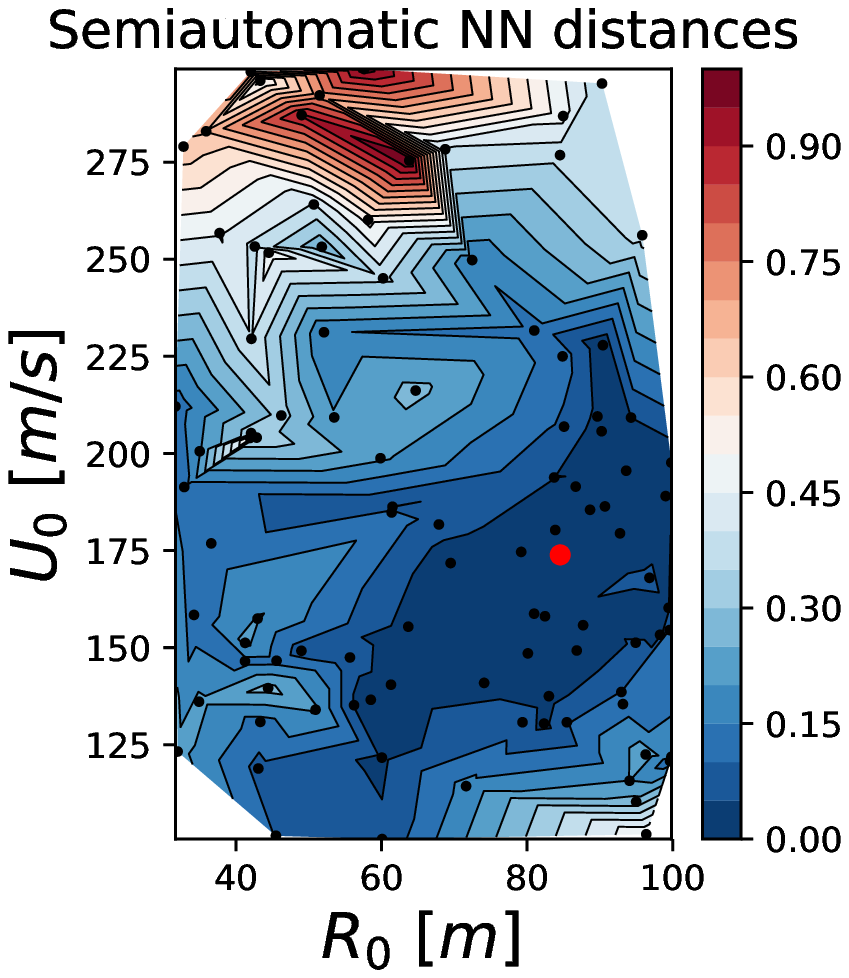}
%				\caption{}
			\end{center}
		\end{subfigure}~
		\begin{subfigure}{0.19\textwidth}
			\begin{center}
				\includegraphics[width=\textwidth]{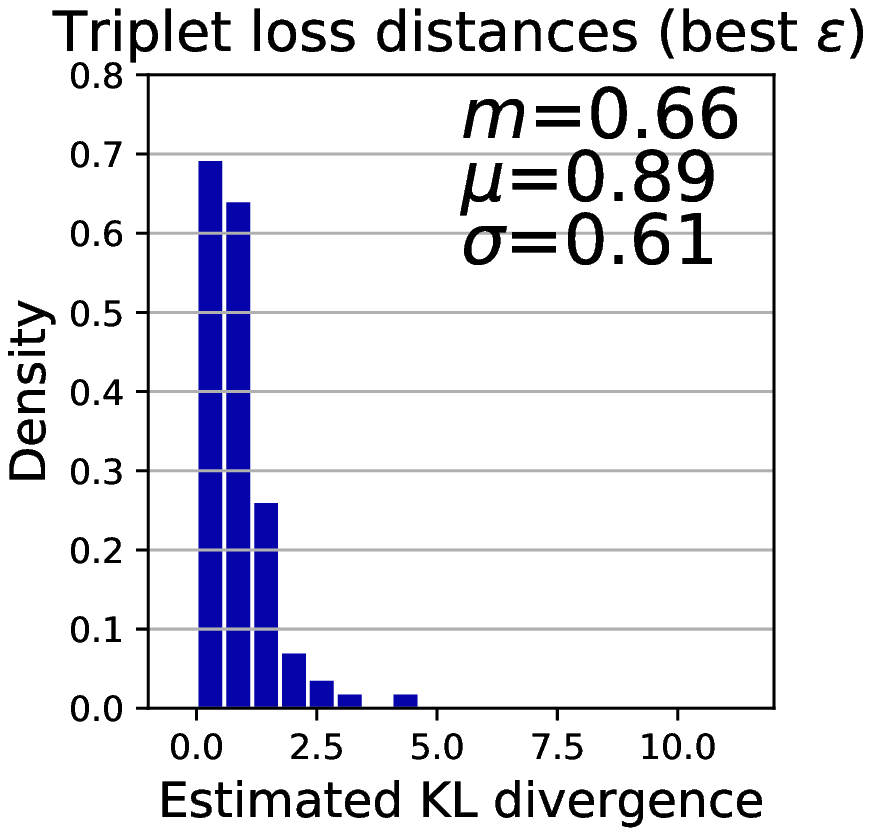}
%				\caption{}
			\end{center}
		\end{subfigure}~
		\begin{subfigure}{0.19\textwidth}
			\begin{center}
				\includegraphics[width=\textwidth]{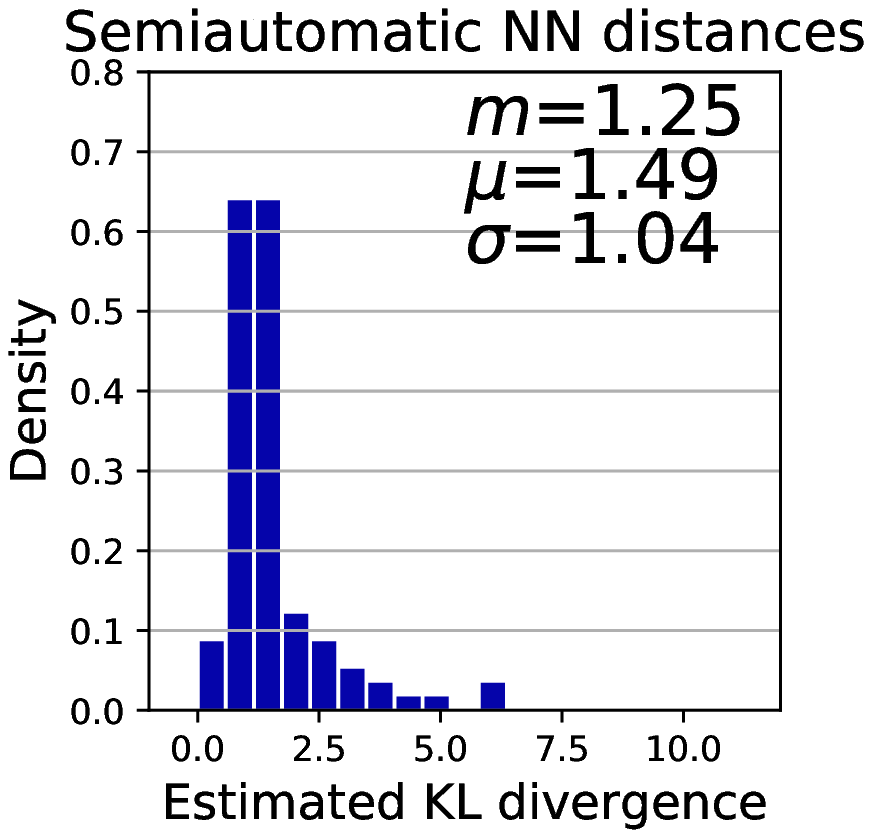}
%				\caption{}
			\end{center}
		\end{subfigure}~
				\begin{subfigure}{0.19\textwidth}
		\begin{center}
			\includegraphics[width=\textwidth]{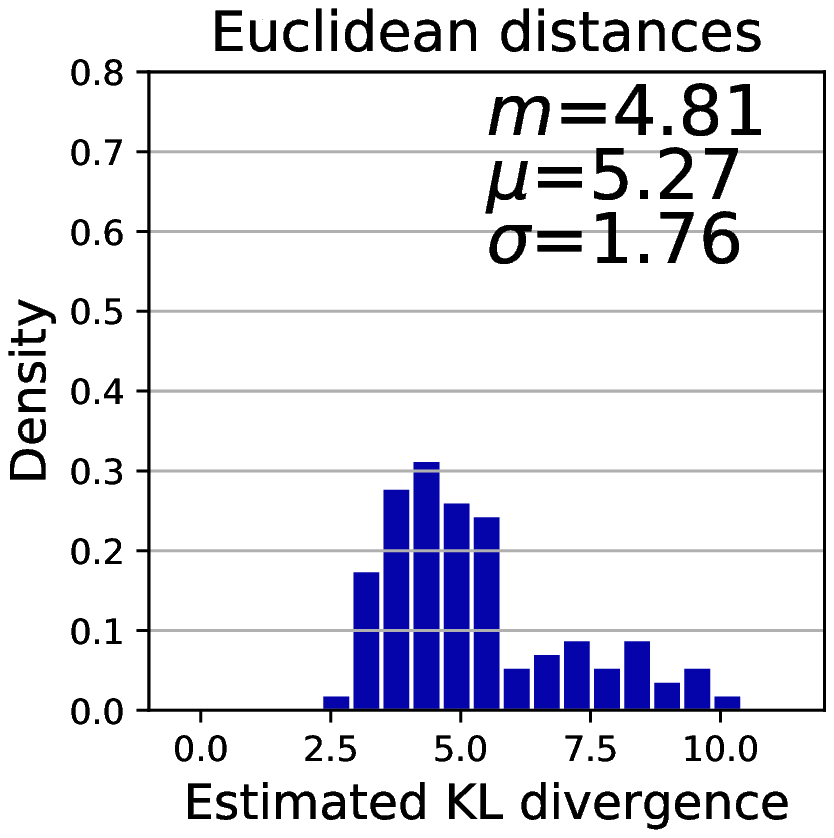}
			%				\caption{}
		\end{center}
	\end{subfigure}
		\caption{\textbf{Comparison between distance learning and summary statistics learning.} We considered the best performing distance learning algorithm according to our analysis, meaning the triplet loss trained over the similarity set corresponding to quantile $ 0.6 $. Although the better performance of triplet loss is not evident from this distance contour plot, it results in being better according to the estimated KL divergence. The inset in the histogram plot reports the median $ m $, the mean $ \mu $ and the standard deviation $ \sigma $ of the estimated KL divergence over the 100 splits.}
		\label{fig:distance_stats_learning_comparison}
	\end{center}
\end{figure}

We compared the performance of this technique with the best distance learning approach that we were able to find, namely the triplet loss one trained over the similarity set defined by using a threshold corresponding to the 60th percentile. In Figure \ref{fig:distance_stats_learning_comparison}, we show both the distance contour plot for the same observation point used in Figure \ref{fig:comp_metric} and the histogram of the estimated KL divergence over the 100 possible splits of the test validation test. For comparison, we also show the histogram of the estimated KL divergence for the Euclidean distance between model outputs. Although the contour plots look very similar, the triplet loss distance is found to slightly outperform the Semiautomatic summary selection technique with neural network according to the KL divergence measure. 

Having demonstrated the stronger performance of the distance learning approach with the triplet loss, we will focus only on that in performing the subsequent ABC inference.

\subsection{Computational considerations}

We stress that ABC with distance or summary statistics learning comes at the expense of a larger computational cost with respect to directly using the Euclidean distance between model outputs in ABC, case in which no training step would be needed. When applying the learning approach, instead, you need to generate the training data, and this is quite expensive given the model we are considering, and then need to perform the training. Note that, once the training data is generated, the SDML technique requires a much shorter time for fitting than the time required for training the neural network in the other cases. However, in the overall balance, when compared with data generation time and the ABC inference time, the training step has a much smaller cost no matter the chosen method, and has to be performed only once. Therefore, it does not make sense to prefer a distance learning method over another just because of a shorter training time. 

During inference, the computation of the new distance only requires multiplying output of the mechanistic model by some matrices with all distance learning techniques (as transforming some data with a neural network simply consists of matrix multiplications and the application of element-wise non-linearity functions). Therefore, the impact of distance learning on the computational complexity of the inference is very small, comparable to the use of hand-chosen summary statistics.

In general, the larger computational cost is balanced by a more efficient ABC inference scheme and a better approximation of the true posterior, given the same computational budget to the inference itself. 
We also remark that our approach can be thought of as a pre-processing technique, as it can be used with any ABC algorithm. Also, once the training has been performed, the same learned distance can be exploited for inference on several observations of the same physical process. 

\subsection{Nested parallelization}
\label{sec:nestparal}
\begin{figure}[htbp]
\begin{center}
\includegraphics[width = \textwidth]{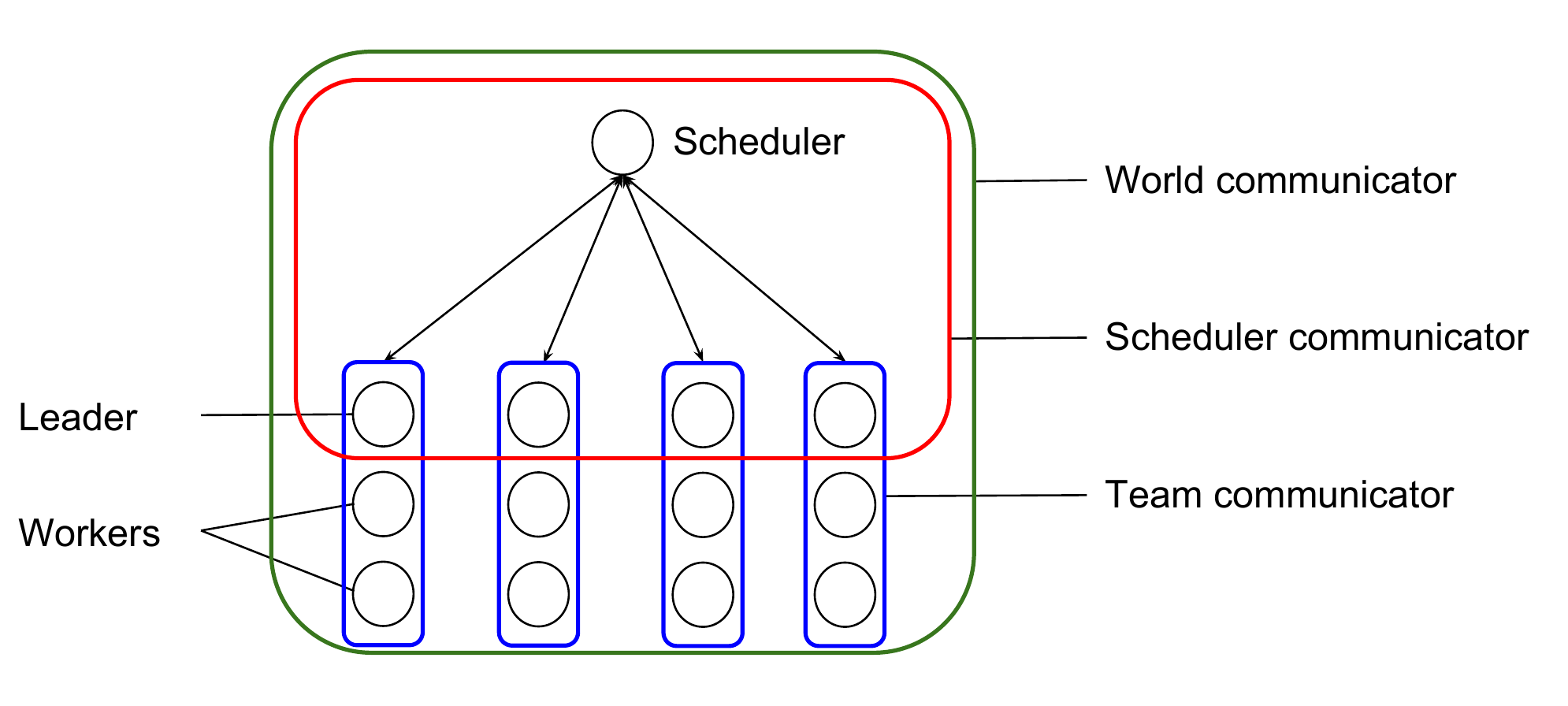}
\end{center}
\caption{\textbf{Nested parallelization:} Description of the communication architecture of the nested MPI parallelization for ABCpy.}
\label{fig:nested_paral}
\end{figure}

For inference, we use the python package ABCpy \citep{Dutta_2017_PASC}, which
implements some of the most advanced ABC algorithms. The algorithms are
implemented such that the computation can be highly parallelized. This is in
particular useful for computationally complex models since the time to solution
for all ABC algorithms is dominated by the models forward simulation time. To be
more precise, the inference algorithms usually need to start forward simulations
with a lot of different model parameters to obtain an accurate posterior
distribution.

ABCpy's backends for parallelization are based on the map-reduce principle. In
the map phase a set of parameters are distributed to a cluster of machines
(nodes) and each node runs forward simulations on the parameters assigned to it.
In the reduce phase the results are collected from the cluster to a single
master node for a next iteration of forward simulations or further processing.
Modern cluster nodes usually have multiple cores and by default ABCpy runs one
forward simulation per core. However, if the model supports multi-threading
(basic operating system threads), the backend can be configured accordingly.

ABCpy provides two different implementations of the map-reduce backend, one
based on Apache Spark and the other based on the message passing interface (MPI)
\citep{MPIForum}. The decision for these technologies was made to cover a broad
user base, since Apache Spark is often used in industry and MPI has its user
base mainly in academia. Nevertheless, MPI has its application beyond
scientific communities in case high throughput and low-latency communication is
required. Sufficiently complex models, as for example in the domains of
meteorology, finite elements, and fluid dynamics, are often parallized using
MPI.

However, previous versions of ABCpy did not support models that were
parallelized using MPI, which is the case for the studied volcano model. 
The challenge in enabling MPI model support is the fact
that MPI code uses an object called MPI communicator to control communication.
In the Apache Spark backend, this communicator is just not available due to the
standard system setup and thus not usable in standard installations. In the MPI
backend, the communicator is available but used by the backend itself that
has to coordinate the parameter distribution and forward simulations.
Thus, we contribute code to ABCpy that enabled support for MPI parallelized models,
broadening the field of applications beyond the volcanic model discussed
here. 

The communication architecture of the nested MPI parallelization is depicted in 
Figure~\ref{fig:nested_paral}. Technically, ABCpy creates two types of 
communicators : The \textit{team communicators} and the 
\textit{scheduler communicator}. Team communicators are used by
the forward simulation models as their main communicator and
one process of each team communicator is part of the scheduler 
communicator. This allows one process, the scheduler, to provides
work to the forward models as long as there are model parameters to explore.

\subsection{Posterior Inference}
\label{sec:postinf}
To draw $Z$ samples approximating the posterior distribution $p(\parameter|\dataObs)$, we keep all the tuning parameters for the APMCABC fixed at the default values suggested in ABCpy package, except the acceptance rate cutoff, which was chosen to be $0.03$. Different number of steps and samples are used for the inference on the simulated and real data; check Section~\ref{sec:result} for more details. 
We consider independent Uniform prior distributions for the parameters with a pre-specified range for each of them, $U_0 \sim U(100,300)\ [m/s]$, $R_0 \sim U(30,100)\ [m]$. 
To explore the parameter space of $\parameter = (U_0, R_0) $, we use a two-dimensional truncated multivariate Gaussian distribution as the perturbation kernel. APMCABC inference scheme centers the perturbation kernel at the sample it is perturbing and updates the variance-covariance matrix of the perturbation kernel based on the samples learned
from the previous step.

For this work, we used supercomputing facilities in the Swiss supercomputing
center (CSCS), namely the Piz Daint supercomputer, where each compute node
consisted of 36 Intel Broadwell Xeon E5-2695 v4 @ 2.10GHz cores. Each forward MPI simulation was run on one node, resulting in the use of $ Z $ cores. With this setup, it is possible to run the APMCABC algorithm with the previously described settings in 2 hours.

\subsection{Parameter estimation}
\label{sec:paramest}
Given an observed dataset $\dataObs$, our main interest is to estimate the corresponding $\parameter$. In decision theory, Bayes estimator minimizes the posterior expected loss, $\E_{p(\parameter|\dataObs)}(\lossfunc(\parameter,\bullet)|\dataObs)$ for an already chosen loss-function $\lossfunc$. If we have $Z$ samples $(\parameter_{i})_{i=1}^{Z}$ from the posterior distribution $p(\parameter|\dataObs)$, the Bayes estimator can be approximated as:
\begin{eqnarray}
\label{eq:Bayes_estimate}
\estparameter_{B}= \argmin_{\parameter} \frac{1}{Z}\sum_{i=1}^Z \lossfunc(\parameter_{i},\parameter). 
\end{eqnarray}
As we consider the Euclidean loss-function $\lossfunc(\parameter,\parameter') = (\parameter-\parameter')^2$, the Bayes estimator can be shown to be the posterior mean $ \E_{p(\parameter|\dataObs)}(\parameter|\dataObs) $, corresponding to an approximate one $\estparameter\approx \frac{1}{Z}\sum_{i=1}^Z \parameter_{i}$.

\section{Results}
\label{sec:result}

\begin{figure}[htbp]
\begin{center}
\includegraphics[width=\textwidth]{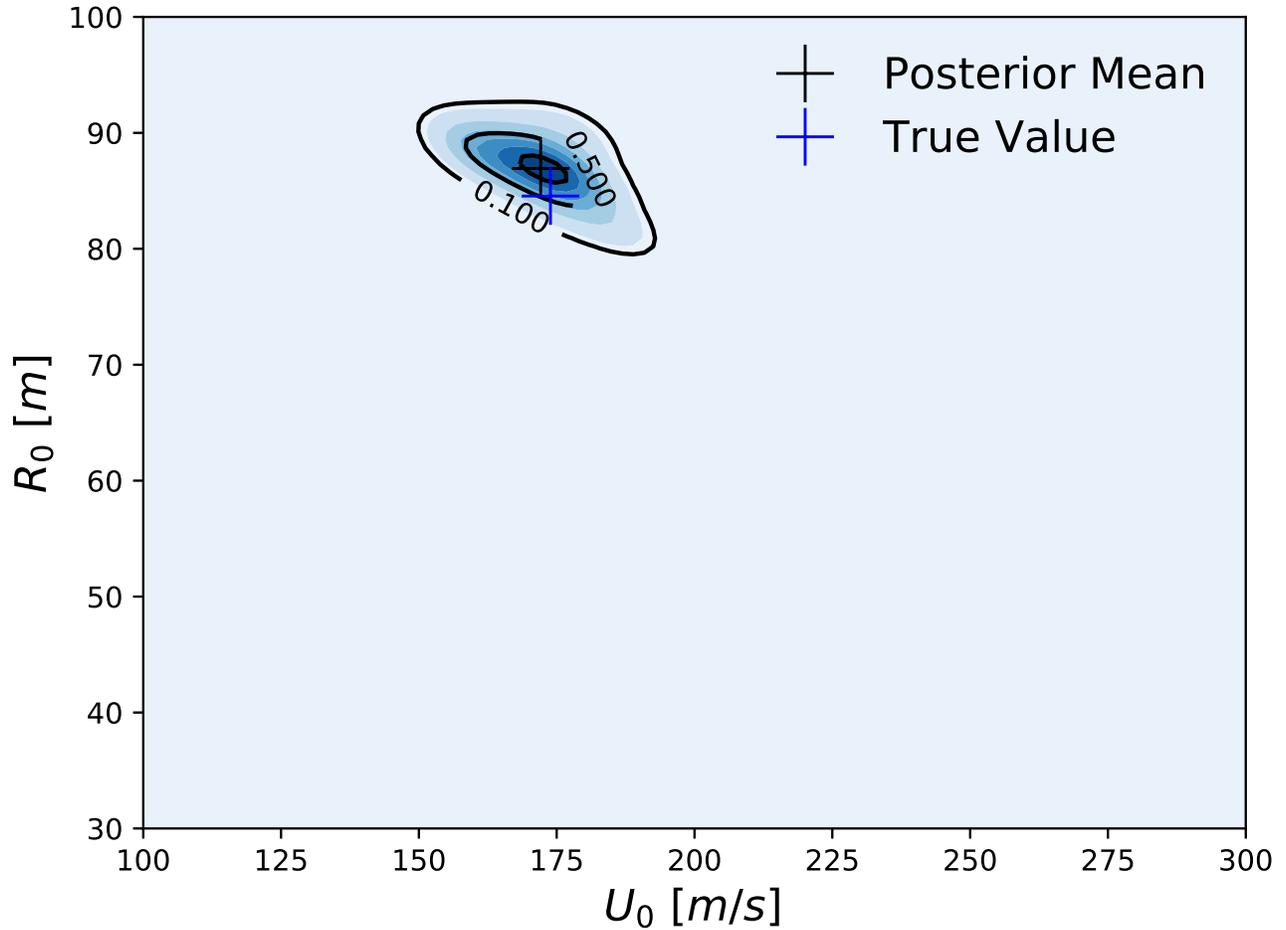}
\caption{\textbf{Inference on Simulated Data:} Approximate posterior contour plot obtained through kernel density (bandwith=0.8) estimate from $Z=100$ samples after 12 iterations of APMCABC scheme from the approximate posterior itself and Bayes estimate (black-cross) of the parameters given a dataset which was simulated from $\Model(\parameter)$ using a known parameter configuration, 
$\parameter^* = (173.87\ m/s, 84.55\ m)$ (blue-cross). The prior on the parameters was chosen to be uniform on the region represented in the plot.}
\label{fig:simul_data}
\end{center}
\end{figure}

To validate the performance of the inference scheme described in Section~\ref{sec:likfinf}, we first try to infer the posterior distribution (using $Z=100$ samples and 12 iterations of APMCABC scheme) and the Bayes estimate of the parameters given a dataset which was simulated from $\Model(\parameter)$ using a known parameter configuration, 
$\parameter^* = (173.87\ m/s, 84.55\ m)$. In Figure~\ref{fig:simul_data}, we plot the approximate posterior distribution inferred by APMCABC using the triple distance learned with best value of $ \epsilon $ according to the above investigation, and the Bayes estimate for $\parameter$. We see the true parameter value $\parameter^* = (173.87\ m/s, 84.55\ m)$ falls in a region with high posterior probability and the Bayes estimate $\hat{\parameter} = (172.09\ m/s, 86.92\ m)$ being close to the true value. 

\begin{figure}[htbp]
\begin{center}
\includegraphics[width=\textwidth]{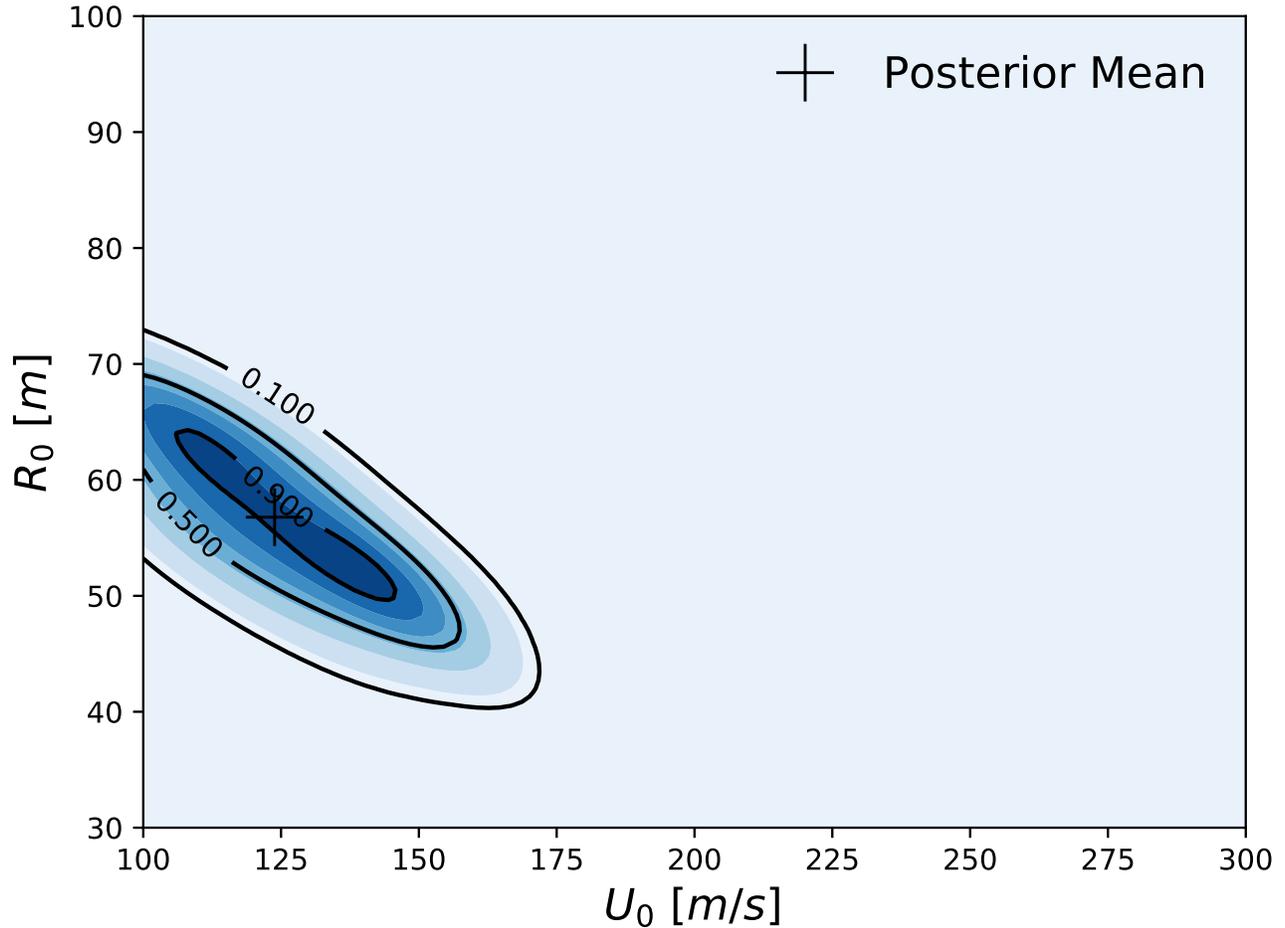}
\caption{\textbf{Inference on Real Data:} Approximate posterior contour plot obtained through kernel density (bandwith=0.8) estimate from $Z=500$ samples after 6 iterations of APMCABC scheme from the approximate posterior itself and Bayes estimate (black-cross) of the parameters given the tephra deposits at 72 ground locations associated with the 2450 BP Pululagua volcanic eruption \citep{volentik2010modeling}. The prior on the parameters was chosen to be uniform on the region represented in the plot.}
\label{fig:real_data}
\end{center}
\end{figure}

After the validation of our inference scheme for a simulated dataset, we perform inference to learn the posterior distribution (using $Z=500$ samples and 6 iterations of APMCABC scheme) and the Bayes estimate of the parameters given the tephra deposits at 72 ground locations associated with the 2450 BP Pululagua (Ecuador) volcanic eruption \citep{volentik2010modeling} in Figure~\ref{fig:real_data}. The Bayes estimate of $\parameter$ and the posterior correlation between $(U_0, R_0)$ are correspondingly $\hat{\parameter} = (\hat{U}_0, \hat{R}_0) = (123.84\ m/s, 56.78\ m)$ and $Corr_{\mbox{post}}(U_0, R_0) = -0.79$, indicating that the similar deposition of tephras could have been caused by combinations of higher injection velocity and narrower vent radius or a lower injection velocity and wider vent radius. 

\begin{figure}[htbp]
\begin{center}
\includegraphics[width=\textwidth]{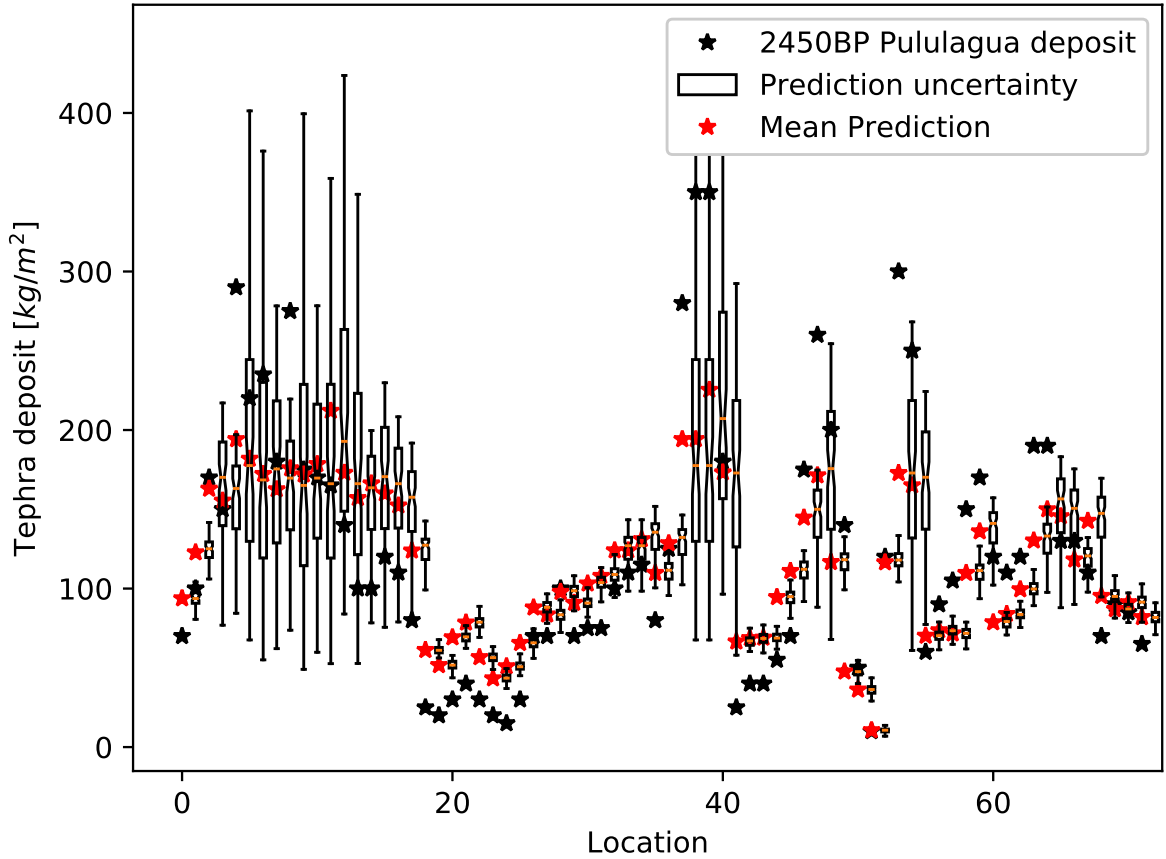}
\caption{\textbf{Posterior Prediction Check:} To validate the numerical volcano model and the inference scheme we perform a posterior prediction check
by simulating 100 datasets, each using a different parameter sample drawn from the posterior distribution. Here, we plot the tephra deposit from field-observation (black) used
for inference, the boxplot of the empirical predictive distribution (white-patch) and the mean predicted dataset (red) at each locations}
\label{fig:prediction_check}
\end{center}
\end{figure}

Next we do a posterior predictive check to validate our model and inference scheme. The main goal here is to analyze the degree to which the observed data deviate from the data generated from the inferred posterior distribution of the parameters. Hence we want to generate data from the model using parameters drawn from the posterior distribution. To do so, we first draw 100 parameter samples from the inferred approximate posterior distribution and simulate 100 data sets, each using a different parameter sample. We call this simulated dataset as the predicted dataset from our inferred posterior distribution and present the mean predicted dataset (red) compared with observed dataset (black) in Figure~\ref{fig:prediction_check}. 
Note that since we are dealing with the posterior distribution, we can also quantify uncertainty in our predictions. We plot the boxplot of the empirical predictive distribution (white-patch) at each locations to get a sense of uncertainty in the prediction. This shows a good prediction performance of the numerical model of volcanic deposition and the proposed inference scheme. The discrepancies in some of the locations can mainly be explained by the inability of the numerical model in providing a complete description of the process of volcanic eruption (eg. the model used in this paper do not consider the variability in deposition due to flow of air). Also, note that some measurement error probably occurred when the data was collected on the field, but that was not reported. However, by taking this into account we would probably be able to reduce the discrepancies. 

\section{Conclusions}
\label{sec:conclusion}
In this manuscript we provide an inferential framework to  calibrate a numerical model of volcanic eruption using ABC. To handle the expensive MPI-parallelized volcanic eruption model, we develop a communication architecture of the nested MPI parallelization for ABC algorithms, implemented in ABCpy. Further we learn a new distance measure to be used in ABC, in a data-driven manner by a deep metric-learning algorithm. 

The present manuscript explored two of the possible techniques for learning a distance exploiting the embedding capabilities of neural networks. Many other techniques have been developed in the computer vision field, and we plan to develop this methodology by adapting more advanced techniques to the ABC setting, in which the training samples do not belong to discrete classes, as it is customary for images, but are only associated to parameter values. Also, we leave the theoretical investigation of the consistency and rate of convergence of the ABC algorithms, using this learned distances, for future work.

In our knowledge, this is the first attempt to use ABC-based inferential schemes for such an expensive stochastic model, which could not be achieved without the development of the nested-parallelization and distance learning. We first validated the inference scheme, by learning the posterior distribution and Bayes estimate for a simulated dataset with a known parameter configuration. Then we applied the inferential framework for a real data consisting of tephra deposits collected in field-observations at 72 ground locations associated with the 2450 BP Pululagua (Ecuador) volcanic eruption \citep{volentik2010modeling}. The developed framework also provides us a framework to quantify prediction uncertainty of this model.  

\section*{Author Contribution} 
Design of the research: RD; Nested Parallelization: PK, MS; Distance-Learning: LP, RD; Writing of  the paper: RD, LP, MS, PK; Contribution to the  writing: BC.

\section*{Code and Dataset}
All the codes used for this article and the datasets can be downloaded from \href{https://github.com/eth-cscs/abcpy-models/tree/master/GeologicalScience/VolcanicEruption}{this repository}.

\section*{Acknowledgements}
We thank CADMOS for providing computing resources at the Swiss Super Computing Center. We also acknowledge partial funding from the European Union Horizon 2020 research and innovation programme for the CompBioMed project (http://www.compbiomed.eu/) under grant agreement 675451.

\bibliographystyle{natbib}

%\bibliography{reference}

\end{document}